\documentclass[lettersize,journal]{IEEEtran}
\usepackage{svg}
\usepackage{amsmath} 
\usepackage{enumitem}
\usepackage{comment}
\usepackage{amssymb} 
\usepackage{soul}
\usepackage{url}
\usepackage{hyperref}
\usepackage{booktabs}
\usepackage{makecell}
\AtBeginDocument{%
  }

\begin{document}

%%
%% The "title" command has an optional parameter,
%% allowing the author to define a "short title" to be used in page headers.
\title{RF-3DGS: Wireless Channel Modeling with Radio Radiance Field and 3D Gaussian Splatting}

%%
%% The "author" command and its associated commands are used to define
%% the authors and their affiliations.
%% Of note is the shared affiliation of the first two authors, and the
%% "authornote" and "authornotemark" commands
%% used to denote shared contribution to the research.

\author{Lihao Zhang, \emph{Student Member, IEEE},  Haijian Sun, \emph{Senior Member, IEEE},  Samuel Berweger, Camillo Gentile, \emph{Member, IEEE}, and Rose Qingyang Hu, \emph{Fellow, IEEE}
\thanks{
L. Zhang and H. Sun (lihao.zhang@uga.edu, hsun@uga.edu) are with the School of Electrical and Computer Engineering, University of Georgia, Athens, GA 30602 USA.

S. Berweger (samuel.berweger@nist.gov) is with the RF Technology Division, National Institute of Standards and Technology, Boulder, CO 80305 USA.

C. Gentile (camillo.gentile@nist.gov) is with the Wireless Networks Division, National Institute of Standards and Technology, Gaithersburg, MD 20899 USA. 

R. Q. Hu (rosehu@vt.edu) is with Bradley Department of Electrical and Computer Engineering, Virginia Tech, Blacksburg, VA 24061 USA. 

Certain commercial equipment, instruments, or materials are identified in this paper in order to specify the experimental procedure adequately. Such identification is not intended to imply recommendation or endorsement by NIST, nor is it intended to imply that the materials or equipment identified are necessarily the best available for the purpose.

}}
\maketitle
%%
%% By default, the full list of authors will be used in the page
%% headers. Often, this list is too long, and will overlap
%% other information printed in the page headers. This command allows
%% the author to define a more concise list
%% of authors' names for this purpose.

%%
%% The abstract is a short summary of the work to be presented in the
%% article.
\begin{abstract} 
Precisely modeling radio propagation in complex environments has been a significant challenge, especially with the advent of 5G and beyond networks, where managing massive antenna arrays demands more detailed information. Traditional methods, such as empirical models and ray tracing, often fall short, either due to insufficient details or because of challenges for real-time applications. Inspired by the newly proposed 3D Gaussian Splatting method in the computer vision domain, which outperforms other methods in reconstructing optical radiance fields, we propose RF-3DGS, a novel approach that enables precise site-specific reconstruction of radio radiance fields from sparse samples. RF-3DGS can render radio spatial spectra at arbitrary positions within 2 ms following a brief 3-minute training period, effectively identifying dominant propagation paths. Furthermore, RF-3DGS can provide fine-grained Spatial Channel State Information (Spatial-CSI) of these paths, including the channel gain, the delay, the angle of arrival (AoA), and the angle of departure (AoD). Our experiments, calibrated through real-world measurements, demonstrate that RF-3DGS not only significantly improves reconstruction quality, training efficiency, and rendering speed compared to state-of-the-art methods, but also holds great potential for supporting wireless communication and advanced applications such as Integrated Sensing and Communication (ISAC). Code and dataset are available at \href{https://github.com/SunLab-UGA/RF-3DGS}{https://github.com/SunLab-UGA/RF-3DGS}.
\end{abstract}

%%
%% The code below is generated by the tool at http://dl.acm.org/ccs.cfm.
%% Please copy and paste the code instead of the example below.

%%
%% Keywords. The author(s) should pick words that accurately describe
%% the work being presented. Separate the keywords with commas.
%% A "teaser" image appears between the author and affiliation
%% information and the body of the document, and typically spans the
%% page.

%%
%% This command processes the author and affiliation and title
%% information and builds the first part of the formatted document.

\section{Introduction}
\subsection{Traditional Wireless Channel Modeling}

Wireless communication systems facilitate the exchange of information carried by Electromagnetic (EM) waves between a physically separated Transmitter (Tx) and Receiver (Rx). As EM waves propagate from the Tx to the Rx, they undergo various effects, including reflection, diffraction, refraction, and scattering. Consequently, these waves may reach the Rx via multiple paths, with each path characterized by its own set of channel properties, known as a multipath component (MPC). 
In the upcoming Sixth-Generation (6G) networks, the antenna array size is expected to grow to thousands of elements under Multiple-Input and Multiple-Output (MIMO). This expansion requires efficient spatial management of such large arrays, thereby mitigating the increasing computational complexity associated with pilot-based estimation as the array size increases~\cite{heath2016MIMOoverview,albreem2021overviewMIMO}. This management, in turn, requires the Spatial Channel State Information (Spatial-CSI)~\cite{Spatial-CSI}, which include not only the traditional path loss and delay of each MPC, but also the angle of departure (AoD), angle of arrival (AoA), polarization and other channel characteristics. Meeting this requirement poses a significant challenge for the wireless channel modeling.

Over the years, various methods have been developed for modeling wireless channel. The empirical models, such as the Okumura-Hata~\cite{okumura-hata2000use} and COST 231 models~\cite{singh2012cost231}, are widely used to predict approximate path loss in large-scale environments. Although effective in providing coarse-grained channel information, they fail to capture the finer details such as the spatial information of the MPCs. 
Computational Electromagnetics (CEM) methods~\cite{bondeson2012computationalEM_book,sullivan2013electromagneticFDTD}, while powerful for small-scale modeling, such as antenna design and near-field communication~\cite{hoorfar2003electromagnetic_antenna,yaghjian1982efficient_nearfield}, are computationally impractical for a wider range of applications. 
Ray tracing, positioned between these two approaches, offers a more appropriate solution by approximating radio waves using ray concepts~\cite{yun2015raytracingradio,charbonnier2020raytracingcalibration}. However, ray tracing is still limited by its high computational complexity and stringent environmental data requirements (e.g., 3D modeling and surface properties), making it unsuitable for more general and real-time applications.

\begin{figure*}[ht]
    \centering
    \includegraphics[width=\textwidth]{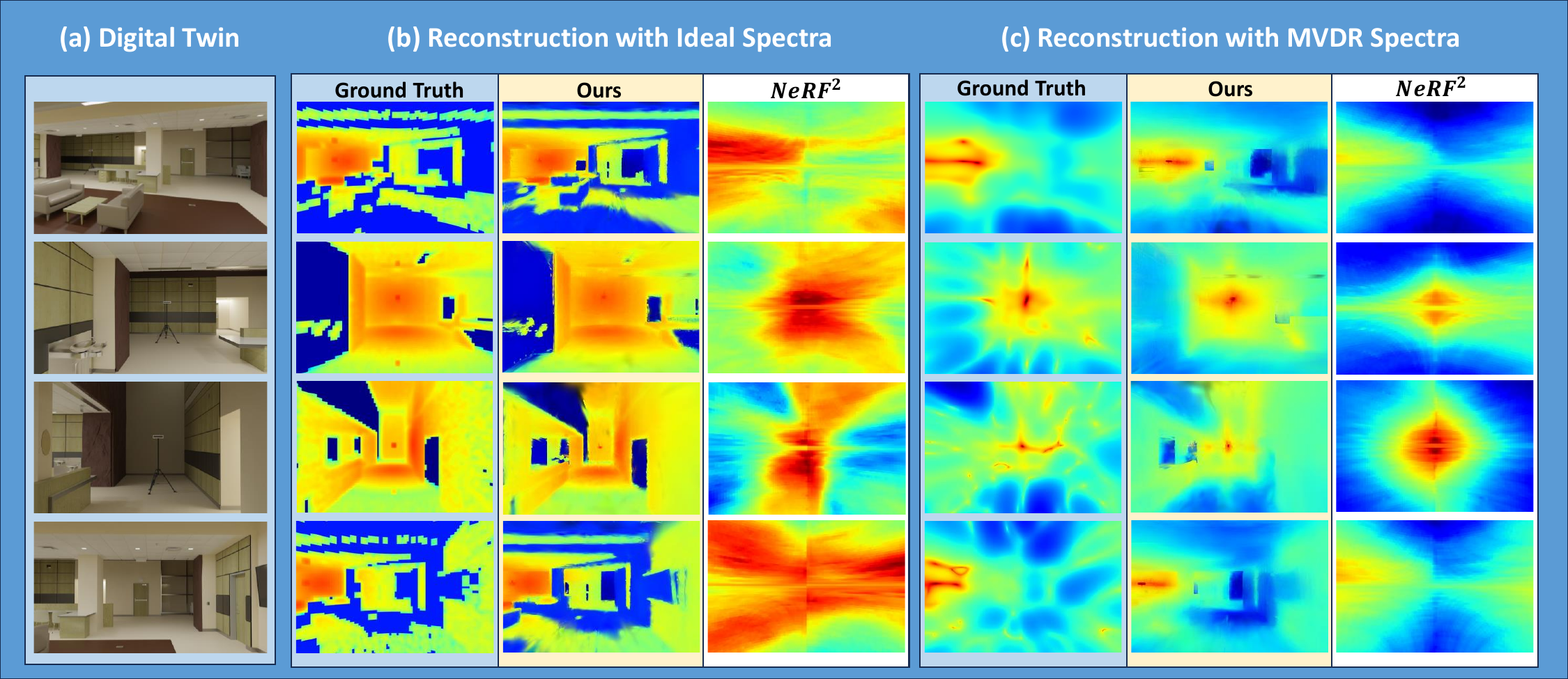} 
    \vspace{-6 mm}
    \caption{Reconstructed radio spatial spectra.\textnormal{ This figure compares the digital twin, the training ground truth, and the spectra reconstructed by our method and \textit{NeRF$^2$}. (a) shows the digital twin's visual photographs. (b) illustrates the group with ideal spectra generated by an ideal array pattern (see Fig~\ref{fig:8_bf_pattern}). (c) presents the group with more practical MVDR spectra generated by an $8\times8$ patch antenna array.}}
    \vspace{-5 mm}
    \label{fig:1_direct_demo}  
   
\end{figure*}

\vspace{-2mm}
\subsection{State-of-the-Art Wireless Channel Modeling}

Recent advancements in wireless channel modeling have introduced a variety of new approaches that leverage environmental information to characterize complex radio propagation with finer granularity.

\textbf{Neural Network Methods:} These approaches utilize rich environmental data as input and employ neural networks to predict channel-related outputs, such as CSI~\cite{linkwiseimai2019radio,kamari2023mmsv} or 2D path loss maps~\cite{levie2021radiounet,2d_bakirtzis2022deepray,3d_raytracingwinert_orekondy2022winert}. While these methods are computationally efficient and yield satisfactory accuracy within the domain defined by their training data, they generally lack determinism and interpretability. For instance, RadioUNet~\cite{levie2021radiounet} employs a UNet architecture to extract multi-scale features from 2D city layouts and infer signal strength across the map. However, its output is limited to 2D path loss values at Rx positions, restricting its applicability to 6G scenarios requiring full Spatial-CSI. Alternatively, mmSV~\cite{kamari2023mmsv} processes 3D street-view imagery to construct a material-labeled 3D scene, which is then fed into a conventional ray tracing engine. This hybrid approach can produce power profiles, but incurs significant overhead due to the need for accurate material segmentation, EM property calibration, and 3D scene alignment.

\textbf{Radiance Field Methods:} 
To model the spatial wireless channel, that is, how EM waves propagate within complex environments, a parallel trend has emerged in the computer vision domain through radiance field methods~\cite{mildenhall2020nerf,muller2022instantnerf}, which reconstruct both geometry and view-dependent surface appearance from sparse optical images. In particular, the view-dependent appearance is modeled as the optical radiance emitted from each surface point. For a given camera, the optical radiance from all visible object points is received as a group of rays arriving at the camera from different pixel directions, similar to how wireless channels are modeled by a group of MPCs arriving at the Rx from different AoAs. 
Building upon this insight, several works have adapted visual radiance field methods to the radio frequency (RF) domain.
A notable example is \textit{NeRF$^2$}~\cite{NeRF2}, which reconstructs a ``squared” radio radiance field by jointly considering both the Tx  and Rx locations as inputs. This framework represents the first attempt to extend visual radiance field modeling to RF propagation, overcoming a key limitation of visual radiance field methods that assume fixed light source positions.

However, \textit{NeRF$^2$} suffers from several limitations. Its reliance on training data generated via Conventional Beamforming (CBF) with small antenna arrays results in severe interference, inter-view inconsistencies, and a lack of geometric fidelity. As a result, achieving high-quality reconstruction requires a dense sampling of spatial views, which undermines the sparse-sampling advantage that NeRF was originally known for. Moreover, inheriting the slow training and inference pipeline of visual-domain NeRF, \textit{NeRF$^2$} is not well suited for efficient RF modeling. Similarly, methods such as RayProNet~\cite{cao2024raypronet}, RFCanvas~\cite{RFcanvas}, and NeWRF~\cite{lu2024newrf} primarily transfer visual radiance field architectures to the RF domain without sufficient adaptation or optimization for radio propagation characteristics.

Regardless of the specific implementation of the radiance field method, reconstructing an effective Radio Radiance Field (RRF) involves three core challenges. First, wireless channels exhibit inherently multi-modal characteristics that differ significantly from optical images, necessitating a tailored representation to capture them. Second, the RRF representation and later reconstruction introduce additional complexity in how to query the representation structure in a differentiable and efficient method. Third, the available training RF data, which are typically noisy and low-resolution radio spatial spectra, pose a challenge for reconstruction.

\subsection{RF-3DGS}

Inspired by the success of 3D Gaussian Splatting (3DGS)~\cite{kerbl3Dgaussians} in visual radiance field reconstruction, we propose RF-3DGS, a novel method that extends 3DGS to the RF domain and addresses the aforementioned challenges. RF-3DGS enables rapid, accurate, and site-specific reconstruction of the RRF using extremely sparse training inputs (as few as 20 for a complex lobby).
Compared to previous RRF methods, RF-3DGS offers several key advantages. In addition to inheriting the strengths of 3DGS, such as ultra-fast training, real-time inference, and compact representation, RF-3DGS introduces two key enhancements: the ability to model multi-modal Spatial-CSI within the RRF, and the capability for super-resolution reconstruction beyond the training radio spatial spectra (i.e., RF ``pictures”). 
These extensions are enabled by two core innovations: the CSI-encoded Spherical Harmonics (SH) function that can capture directional Spatial-CSI, and the two-stage fusion training strategy, which integrates visual information to support the reconstruction process. 

To support and validate RF-3DGS, we construct a digital twin framework featuring a high-fidelity digital replica and co-located field measurements. Multiple array signal processing (ASP) methods are applied to generate diverse training spectra, enabling comparative evaluation with RF-3DGS and baseline methods.

Fig.~\ref{fig:1_direct_demo} presents visual comparisons under both ideal and practical Minimum Variance Distortionless Response (MVDR) spectra. RF-3DGS consistently reconstructs accurate RRFs in unseen areas with sparse input and exhibits strong extrapolation enabled by fusion training. Quantitatively, it reduces reconstruction error by 84.64\% over \textit{NeRF$^2$}, with a training time of 3 minutes and inference speed of 2 ms—compared to 3 hours and 1 s for \textit{NeRF$^2$}.
We further demonstrate RF-3DGS’s utility in wireless communication by using the synthesized spectra to guide CBF beamforming. In the ideal case, RF-3DGS achieves a median steering error of 5.94$^{\circ}$. Additionally, reconstructed AoD and delay spectra align closely with ground truth. Comparison with physical twin measurements confirms the fidelity of our digital twin setup.

\subsection{Contributions}
Our main contributions are as follows:
\begin{itemize}[left=0pt]
    \item We formulate the theoretical foundations of RRF modeling, including its representation structure, query mechanism, and reconstruction from sparse wireless samples.
    
    \item We introduce RF-3DGS, a novel framework that encodes multi-modal channel characteristics via CSI-encoded Spherical Harmonic functions, and fuses visual and radio data through a two-stage training strategy to overcome practical limitations.
    
    \item We develop a digital twin platform combining real-world measurements and ray-traced simulations, and demonstrate the benefits of RF-3DGS for massive MIMO spatial management and Integrated Sensing and Communications (ISAC) applications.
\end{itemize}

\begin{table}[h]
\caption{Key Terms and Notations}
\vspace{-5mm}
\begin{center}
\begin{tabular}{c|p{5.5cm}}
\hline
\textbf{Term} & \textbf{Description} \\
\hline
Antenna Array & A 2D uniform planar array (UPA) or other configurations with sufficient elements to control the 3D array gain pattern for transmission and reception. \\
\hline
Spatial-CSI & Spatial characteristics of a multi-path channel between a Tx–Rx pair. \\
\hline
Spatial-CSI Value & A vector representing the spatial characteristics of a specific MPC. (This paper will focus on path loss, normalized delay, AoA, and AoD.) \\
\hline
Pose & Position and orientation of an array or camera in world coordinates. \\
\hline
$\mathbf{x}, \mathbf{x}_{\text{obj}}, \mathbf{x}_{\text{Rx}}$ & A generic 3D point, an object position, and a receiver position, respectively. \\
\hline
$\mathbf{d}$, $\mathbf{d}_{\text{out}}$, $\mathbf{d}_{\text{query}}$ & A unit 3D direction vector; the outgoing direction from object to Rx; and the query direction from Rx to object, respectively. \\
\hline
$\mathbf{c}(\mathbf{x}, \mathbf{d})$ & Vector-valued RRF function over all positions $\mathbf{x}$ and outgoing directions $\mathbf{d}$. \\
\hline
$\mathbf{c}_{\mathbf{x}}(\mathbf{d})$ & Vector-valued spherical function describing radiance at a given point $\mathbf{x}$ over all outgoing directions $\mathbf{d}$. \\
\hline
$\mathbf{c}_{\mathbf{x}, \mathbf{d}}$ & RRF function vector output for a given position–direction pair $(\mathbf{x}, \mathbf{d})$. \\
\hline
$\alpha(\mathbf{x})$ & A scalar-valued density field function. \\
\hline
$\mathbf{m}(\mathbf{x}_{\text{Rx}}, \mathbf{d}_{\text{query}})$ & Spatial-CSI value of the MPC queried at receiver position $\mathbf{x}_{\text{Rx}}$ on direction $\mathbf{d}_{\text{query}}$. \\
\hline
\end{tabular}
\label{tab:notation_ieee}
\end{center}
\vspace{-8mm}
\end{table}

\section{Fundamentals of the RRF Method}
\label{sec:2}

Before detailing RF-3DGS, we first establish the theoretical foundations of the Radio Radiance Field method. This begins by describing how the traditional multipath wireless channel model can be represented using the RRF. To avoid confusion caused by prematurely introducing RF-3DGS, given its multiple advanced techniques, we first present a straightforward but naive RRF reconstruction method to offer a global view of how RRF-based modeling works.

\subsection{From Deterministic Multipath Model to Radio Radiance Field}

\begin{figure}[h]
    \vspace{-3mm}
    \centering
    \includegraphics[width=0.45\textwidth]{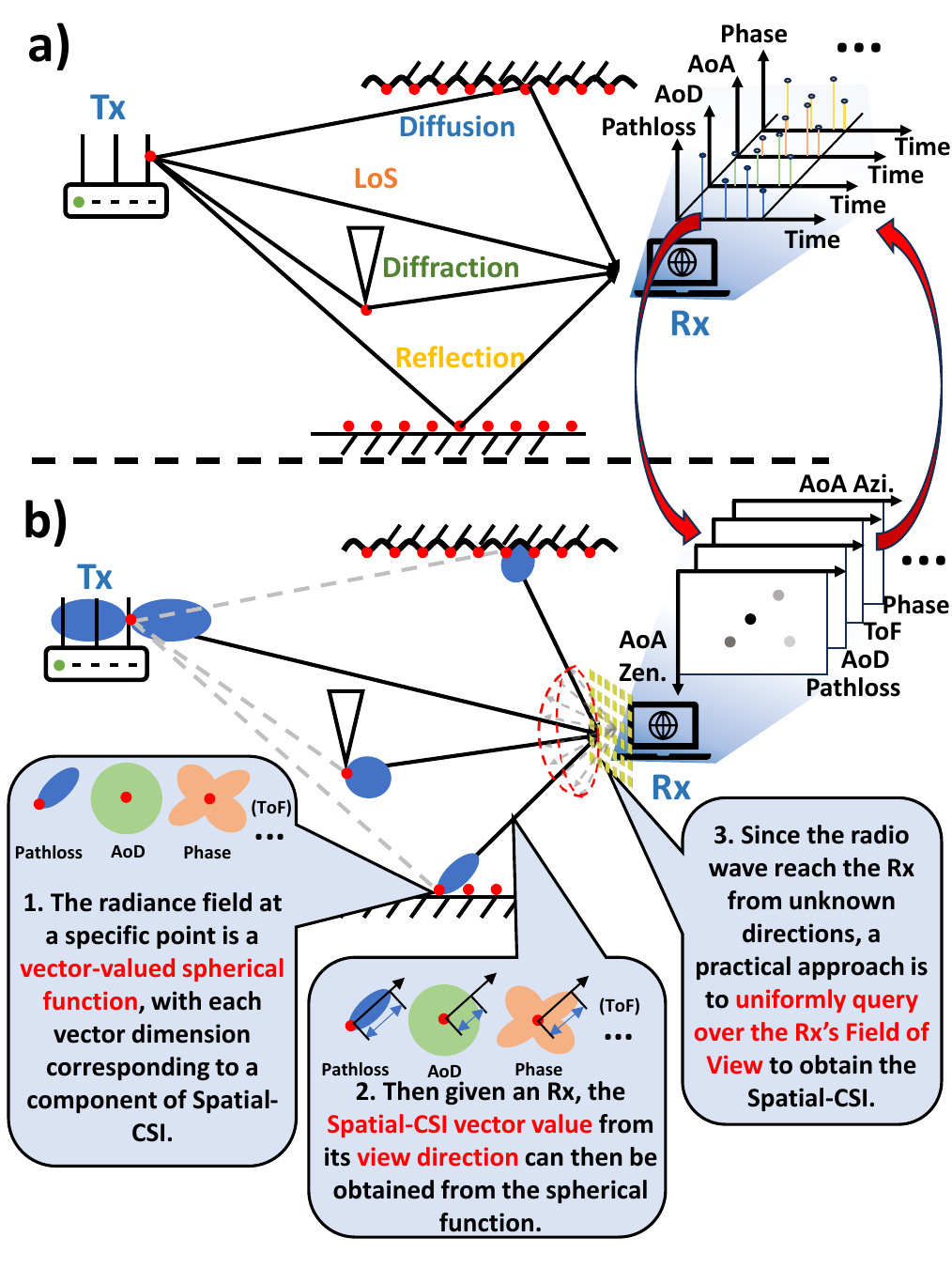}
    \vspace{-3mm}
    \caption{From deterministic multipath model to radio radiance field.}
    \label{fig::2_Radio Radiance Field} 
    \vspace{-1mm}
\end{figure}

For a well-understood environment and a specific Tx-Rx pair within it, the wireless channel can be deterministically modeled as a set of rays originating from the Tx, propagating along different paths, and reaching the Rx as distinct MPCs with unique Spatial-CSI values, as the four paths shown in Fig.~\ref{fig::2_Radio Radiance Field}(a).
Such a deterministic multipath model is typically visualized using multi-dimensional Channel Impulse Response (CIR) plots. In such plots, each impulse corresponds to an MPC, the horizontal axis indicates the arrival time of the MPCs, and the vertical axis represents Spatial-CSI components, such as path loss, AoD, AoA, phase, and others, as shown in the right side of Fig.~\ref{fig::2_Radio Radiance Field}(a). To comprehensively represent such spatial information across an environment, conventional approaches often densely sample and store these CIRs over a 3D grid of Rx positions, leading to massive measurement efforts, high processing demands, and storage costs in the tens of gigabytes~\cite{wisegrt,NIST_data}. The need for high-fidelity interpolation between discrete points further adds complexity.
In contrast, RRF offers a more efficient, physics-consistent, and scalable alternative. Rather than storing channel responses at each Rx point, RRF models how radio radiance emerges from object surfaces, encapsulating the most compact underlying physical interactions that govern radio propagation. In the following, we first show how the MPC of an arbitrary path can be represented using the RRF. Then, we demonstrate how the full multipath model at an arbitrary Rx location can be derived from the RRF representation.

First and foremost, the core of the RRF concept is the vector-valued RRF function $\mathbf{c}(\mathbf{x}, \mathbf{d})$, a field function. In its general form, this function is defined over all continuous 3D positions $\mathbf{x}$ and for each position, describes the radio radiance emitted over all continuous directions $\mathbf{d}$. However, a common misconception is to treat the RRF as traditional field concepts such as the EM field. In an EM field, once a source exists, almost all points in space, including those free space points, possess a well-defined field vector. However, the RRF fundamentally differs in that it only describes radiance sources. For example, although a free space point may be passed through by certain radio waves, this point does not emit any radiance and hence is not considered a source. Therefore, the RRF value at such a point is zero in all directions. Based on this understanding, a more practical formulation of the RRF function focuses solely on object points, denoted as $\mathbf{c}(\mathbf{x}_{\text{obj}}, \mathbf{d})$. (While not all object points necessarily act as radiance sources, it is both general and convenient to assume that radiance may originate from all such points.)

As shown in Fig.~\ref{fig::2_Radio Radiance Field}(b.1), at the pointed surface point $\mathbf{x}_{\text{obj}}$ corresponding to the reflection path, the RRF function $\mathbf{c}(\mathbf{x_{\text{obj}}}, \mathbf{d})$ reduces to a vector-valued spherical function $\mathbf{c}_{\mathbf{x}_{\text{obj}}}(\mathbf{d})$. Each dimension of this $\mathbf{c}_{\mathbf{x}_{\text{obj}}}(\mathbf{d})$ represents the distribution of a Spatial-CSI component (e.g., path loss, AoD, phase, and so on) across all directions $\mathbf{d}$. 
If an Rx is in line-of-sight (LoS) with this point, we can get the outgoing direction $\mathbf{d}_{\text{out}}$ from this point to the Rx, and assume the radio radiance from this point is received as an MPC at the Rx. For this MPC, the vector Spatial-CSI value can be extracted as $\mathbf{c}_{\mathbf{x}_{\text{obj}},\mathbf{d_{\text{out}}}}$, as shown in Fig.~\ref{fig::2_Radio Radiance Field}(b.2). This MPC corresponds to the reflection path in Fig.~\ref{fig::2_Radio Radiance Field}(a). Since the RRF function describes all potential radiance sources and outgoing directions, we have proved that the MPC of an arbitrary radio propagation path can be modeled using the RRF function.

The remaining question is whether we can recover the full multipath channel at an arbitrary Rx position $\mathbf{x}_{\text{Rx}}$. For such an Rx location, the AoAs of the dominant MPCs are unknown. Therefore, as illustrated in Fig.~\ref{fig::2_Radio Radiance Field}(b.3), a practical strategy is to uniformly query a set of discrete directions $\mathbf{d}_{\text{query}}$ within the Rx’s Field of View (FoV) (i.e., the angular range over which the Rx maintains valid gain), to capture the incoming MPCs.
On each query direction $\mathbf{d}_{\text{query}}$, a ray is cast to identify the first intersected point $\mathbf{x}_{\text{obj}}$. Assuming there exists an MPC outgoing from this point in the direction $-\mathbf{d}_{\text{query}}$, its corresponding Spatial-CSI value can be retrieved from the RRF as $\mathbf{c}(\mathbf{x}_{\text{obj}}, -\mathbf{d}_{\text{query}})$. After performing such queries across all directions, the results can be organized into a multi-dimensional radio spatial spectrum, or an ``RF picture,” as shown in the upper right of Fig.~\ref{fig::2_Radio Radiance Field}(b). Similar to an RGB image, this RF picture has multiple channels, and the queried Spatial-CSI values of the MPCs become its pixel values. The 2D coordinates of each pixel (i.e., the row and column indices) correspond to the azimuth and zenith angles of the AoA of the MPC. In Fig.~\ref{fig::2_Radio Radiance Field}(b), the example RF picture contains four distinct points, corresponding to the four propagation paths shown in Fig.~\ref{fig::2_Radio Radiance Field}(a). (In Fig.~\ref{fig:1_direct_demo}, the spectra contain many more weakly diffused MPCs, making them more picture-like.)
To this end, we can say that the radio spatial spectrum (RF picture) obtained from the RRF is equivalent to the traditional multipath model represented by the CIR. The difference lies in the domain of unfolding: the CIR unfolds MPCs in the time domain, while the RRF unfolds them in the AoA domain. %However, as will be shown later, the RRF offers a more compact, physically consistent, and training-data-efficient method for reconstructing radio propagation in 3D space.

\vspace{-2mm}

\subsection{A Simple RRF Reconstruction Framework}
To better understand the technical design of RF-3DGS, we first introduce a straightforward but naive RRF reconstruction method. This includes how to represent the RRF using a learnable structure, how to perform differentiable queries of Spatial-CSI along each ray, and how to reconstruct the RRF from training data.

\textbf{Learnable Representation:} First, modeling the RRF function alone is insufficient in practice, as the environment geometry is also essential. For a query ray emitted from the Rx, it is necessary to determine which object surface point the ray intersects first in order to extract its radiance. Therefore, a practical RRF representation must include both the RRF function and the scene geometry.
To model geometry, we adopt one of the simplest and most general formulations: the density field function $\alpha(\mathbf{x})$. This function assigns values to all continuous 3D positions $\mathbf{x}$: it is zero for free space, low for translucent objects, and high for solid objects. Then, we can define a practical RRF representation as $ \mathcal{R}(\mathcal{E}, \mathcal{T}) = \left\{ \mathbf{c}(\mathbf{x_{\text{obj}}},\mathbf{d}),\ \alpha(\mathbf{x}) \right\}$. This RRF representation $\mathcal{R}$ maps the wireless communication scenario (given the environment $\mathcal{E}$ and the Tx $\mathcal{T}$) to the corresponding RRF function $\mathbf{c}(\mathbf{x_{\text{obj}}}, \mathbf{d})$ and density field function $\alpha(\mathbf{x})$.
To make the representation learnable, we use two Multi-Layer Perceptrons (MLPs) to learn the $\mathbf{c}(\mathbf{x_{\text{obj}}}, \mathbf{d})$ and $\alpha(\mathbf{x})$, respectively.

\begin{figure}[h]
    \vspace{-3mm}
    \centering
    \includegraphics[width=0.35\textwidth]{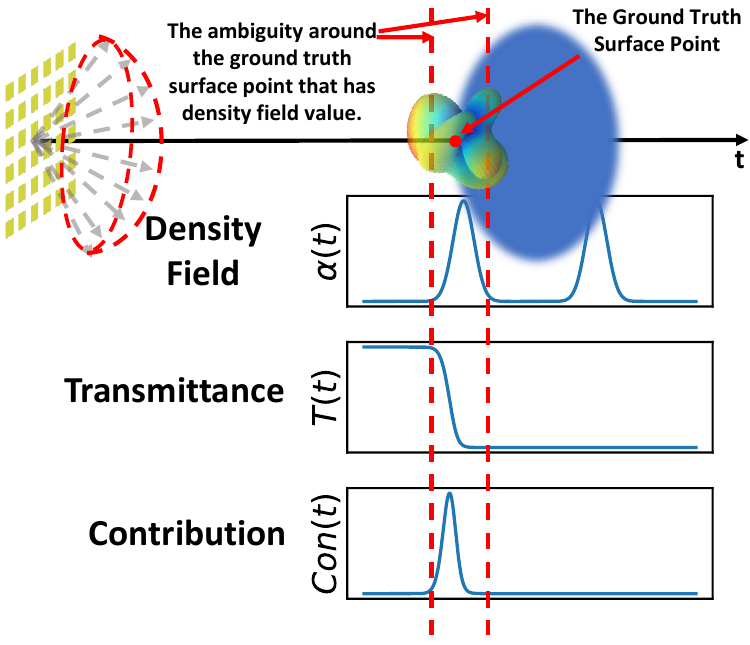}
    \vspace{-3mm}
    \caption{Alpha-blending on a ray in the RRF representation.} 
    \vspace{-2mm}
    \label{fig::3_RRF rendering} 
\end{figure}

\textbf{Differentiable Query:}  
To obtain the multipath model at a given Rx location, we previously discussed querying a set of directions within the FoV. The remaining challenge is how to retrieve the Spatial-CSI value along each ray in a differentiable manner.
In the ideal case, assuming a scene composed solely of solid objects and a perfectly trained geometry MLP that models extremely thin object surfaces, the problem is solved as just identifying the first intersected point along a query ray, as illustrated by the red ground-truth surface point in Fig.~\ref{fig::3_RRF rendering}. 

However, in practice, the geometry MLP (as well as most RRF methods) often struggles to represent thin surfaces, leading to ambiguity in the learned density field function near the ground truth object surface, as illustrated by the region between the two red dotted lines and the corresponding density field curve in Fig.~\ref{fig::3_RRF rendering}. This necessitates a query strategy that is adaptive to such ambiguity. Another reason for such adaptability is that, during reconstruction, the geometry representation is also gradually optimized from a loose approximation to a dense geometry, and the querying method must accommodate this gradual densification. Furthermore, it should also handle translucent objects appropriately, where rays should not immediately terminate upon intersecting a semi-transparent medium. (It is worth noting that the learned density field only captures surface geometry, as internal structures cannot be inferred from either visual or RF observations. This explains the empty interior region in the illustrated density curve.)

Therefore, for this adaptability, we assign varying contributions to different regions along a query ray, considering two key factors. First, density: regions with higher density along the ray should contribute more to the final radiance. Second, occlusion: closer regions with nonzero density should attenuate the contributions from farther regions.  
To incorporate these factors in a differentiable and physically plausible manner, we adopt the alpha-blending technique. Formally, for a receiver position $\mathbf{x}_{\text{Rx}}$ and a query direction $\mathbf{d}_{\text{query}}$, the queried Spatial-CSI value $\mathbf{m}(\mathbf{x}_{\text{Rx}}, \mathbf{d}_{\text{query}})$ is defined as:
\begin{equation}
\mathbf{m}(\mathbf{x}_{\text{Rx}}, \mathbf{d}_{\text{query}}) = \int_{t_n}^{t_f} \mathbf{c}(t,- \mathbf{d}_{\text{query}}) \cdot T(t) \cdot \alpha(t)\, dt,
\label{eq::volume_rendering_eq}
\end{equation}
\begin{equation}
T(t) = \exp\left(-\int_{t_n}^{t} \alpha(s)\, ds\right),
\label{eq::transmittance_eq}
\end{equation}
where t is the 1D coordinate on the ray, $T(t)$ denotes the transmittance along the ray from the near plane $t_n$ to a point $t$, $\alpha(t)$ is the density field function value, and $t_f$ is the predefined upper bound of integration. Specifically, the radiance $\mathbf{c}(t,- \mathbf{d}_{\text{query}})$ of a point $t$ along the ray is weighted by both its density $\alpha(t)$ and its transmittance $T(t)$. The transmittance $T(t)$ is computed as the accumulated density in front of point $t$, reflecting the occlusion relationship along the ray. Consequently, as illustrated in the lower-right plot of Fig.~\ref{fig::3_RRF rendering}, this formulation ensures that only regions near the ground truth surface contribute significantly to the final result.
Although Equations~(\ref{eq::volume_rendering_eq}) and ~(\ref{eq::transmittance_eq}) describe the continuous formulation of alpha-blending, in practice, MLPs can only infer values at discrete points along the ray. Therefore, a discretized version of alpha-blending must be used, which requires additional sampling strategies to balance accuracy and computational efficiency.  
The detailed formulation of such discrete querying is beyond the scope of this paper, as RF-3DGS adopts an advanced differentiable querying method that avoids the need for discretization.

\begin{figure}[h]
    \centering
    \includegraphics[width=0.5\textwidth]{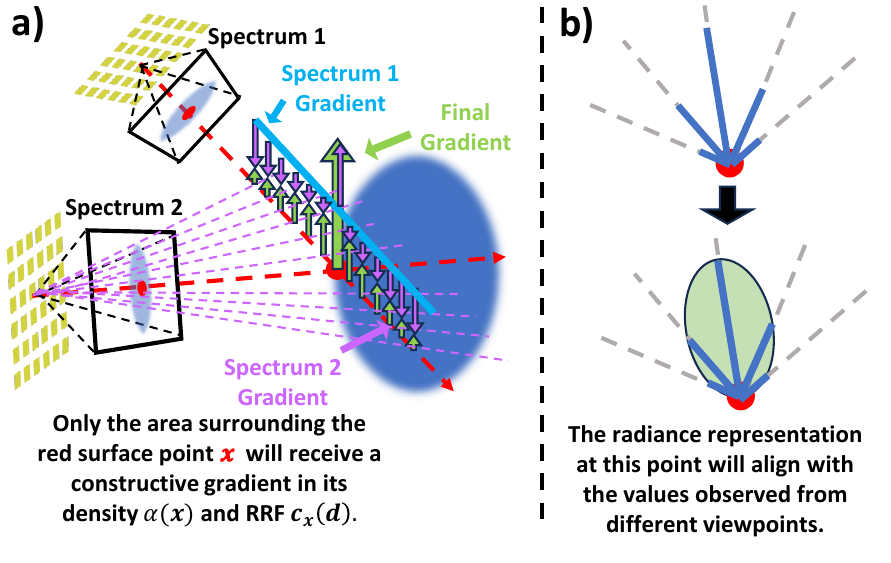}
    \vspace{-8mm}
    \caption{Reconstruct RRF from radio spatial spectra.} 

    \label{fig::4_RRF reconstruction} 
\end{figure}

\begin{figure*}[ht]

    \centering
    \includegraphics[width=1.0\textwidth]{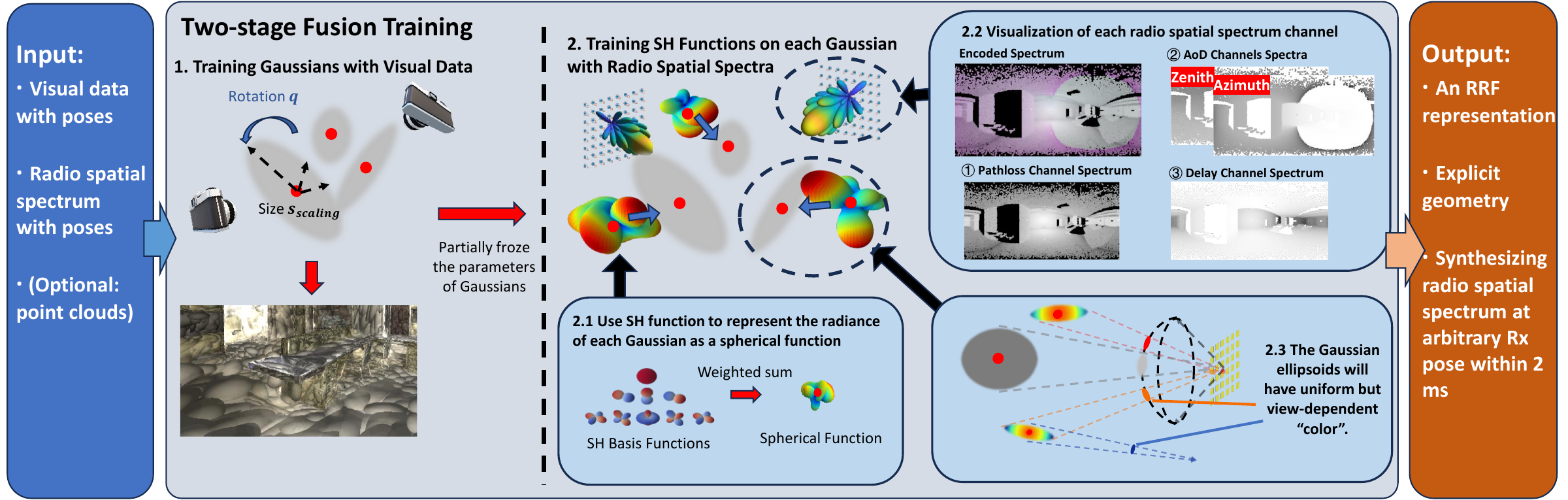}
 
    \caption{RF-3DGS pipeline.} %\textnormal{(a) The extracted geometry representation from the visual data, consisting of millions of Gaussians, where the Gaussians are visualized by their contour surfaces. (b) The reconstructed radio radiance field, where the objects are illuminated by RF waves. (c) A demonstration of SH functions. (d) An example of the CSI-encoded spectrum and the corresponding decoded CSI spectra from the respective channels.}}

    \label{fig:5-RF-3DGS pipeline}
    \vspace{-3mm}
\end{figure*}

\textbf{Reconstruction:}  
Given the learnable MLP representations, and the differentiable alpha-blending querying, the reconstruction can be naturally performed through gradient descent. During training, the queried Spatial-CSI value along a ray from the MLP is compared to the corresponding ground-truth value, and the resulting loss is backpropagated to update the MLPs. However, a key question arises: how does the optimizer ensure that only the true surface region along a ray receives constructive gradients, given that alpha-blending integrates contributions without explicitly considering spatial location? The answer lies in the training from different Rx poses.

As illustrated in Fig.~\ref{fig::4_RRF reconstruction}(a), two training radio spatial spectra are captured from different Rx poses but observe the same red surface point. Consequently, the two red query rays, which belong to the two spectra respectively, intersect at this point. 
At the beginning of training (zero initialization), we assume the iteration starts from spectrum 1. On its red query ray, the gradients are distributed uniformly along the ray due to the absence of density field prior, as indicated by the blue line. As training proceeds to spectrum 2, multiple purple query rays intersect, or pass very close to, the same region traversed by the red ray in spectrum 1. These overlapping purple rays lead to the correction of previously misallocated gradients from spectrum 1, as illustrated by the purple arrows.
In this process, only the region near the true surface point still receives constructive gradient updates. The combination of gradients from spectrum 1 and spectrum 2 ultimately produces a concentrated gradient around the correct surface point, represented by the green arrows on the red query ray from spectrum 1.

In later training iterations, this region, now with higher learned density, will contribute more to the alpha-blending integral (due to increased $\alpha(t) \cdot T(t)$), thereby attracting more constructive gradient. This positive feedback loop accelerates convergence of the density field toward the true surface geometry.
Regarding the radiance representation, as shown in Fig.~\ref{fig::4_RRF reconstruction}(b), a surface point is observed from multiple Rx poses, each associated with a corresponding ground-truth Spatial-CSI vector. This allows the radiance MLP to gradually fit the correct directional pattern through the supervision from the multiple Rx poses.

\section{Proposed RF-3DGS Framework}
Although we have introduced a naive RRF reconstruction pipeline and provided an intuition about how such methods function, three critical challenges remain for achieving efficient and high-quality RRF reconstruction. 
First, designing a more efficient representation is non-trivial. Existing NeRF-based RRF methods rely on MLPs, which are redundant, slow to train, and inherently implicit—making it difficult to directly extract usable information for downstream tasks. Second, efficient rendering and training strategies are required to improve both reconstruction speed and quality. Existing implementations based on MLPs, voxel grids, or tensor decompositions often depend on complex sampling strategies, which substantially hinder their computational efficiency and scalability. Third, there is an inherent lack of geometric information in the radio spatial spectra. As illustrated in Fig.~\ref{fig:1_direct_demo}, even the ideal radio spectra exhibit low resolution and noticeable distortion, making them insufficient for precise RRF reconstruction.

\vspace{-3mm}
\subsection{Geometry Representation with 3D Gaussian Distributions}
Unlike NeRF, which uses two MLPs to directly fit the density field function $\alpha(\mathbf{x})$ and the RRF function $\mathbf{c}(\mathbf{x}, \mathbf{d})$ for all continuous points $\mathbf{x}$, 3DGS utilizes millions of learnable 3D Gaussian distributions (hereafter referred to as ``Gaussians'') as the geometry representation. In RF-3DGS, we adopt a similar configuration, using Gaussians as the basic primitives for the geometry. Each Gaussian has its own learnable parameters, including the basic density $\alpha_g$, the center (mean) of the Gaussian $\mathbf{x}$, and the shape of the Gaussian, defined by a $3 \times 3$ covariance matrix $\boldsymbol{\Sigma}$. The density field of a single 3D Gaussian distribution can be represented as:
\begin{equation}
    \alpha(\mathbf{x}) = \alpha_g \cdot e^{-\frac{1}{2} (\mathbf{x})^\mathrm{T} \boldsymbol{\Sigma}^{-1} (\mathbf{x})}
\end{equation}
However, due to the fact that the covariance matrix of a Gaussian must be positive semi-definite, it cannot be directly constrained during gradient descent. Therefore, the shape of the 3D Gaussians is represented by a more constrained set of parameters. The covariance matrix can be decomposed as:
\begin{equation}
    \boldsymbol{\Sigma} = \mathbf{R} \mathbf{S} \mathbf{S}^\mathrm{T} \mathbf{R}^\mathrm{T}
\end{equation}
where $\mathbf{S}$ is a $3 \times 3$ scaling diagonal matrix, and its diagonal elements constitute a 3D scaling vector $\mathbf{s}_{scal}$ that controls how the 3D Gaussian spreads along the three dimensions. The matrix $\mathbf{R}$ is a $3 \times 3$ rotation matrix (an orthogonal matrix), which can be mapped with a quaternion $\mathbf{q}$ that controls the rotation of the Gaussian, as shown in Fig.~\ref{fig:5-RF-3DGS pipeline}. Both $\mathbf{s}_{scal}$ and $\mathbf{q}$ are easy to constrain within valid domains. For instance, large, anomalous Gaussians with excessive $\mathbf{s}_{scal}$ values can be excluded, and all $\mathbf{q}$ values can be normalized to unit length, ensuring validity.

Thus, during the training process, the parameters of the millions of Gaussians are optimized, ``moving" and ``transforming" these Gaussians to precisely represent the geometry density field $\alpha(\mathbf{x})$. This is illustrated in Fig.~\ref{fig:5-RF-3DGS pipeline}, where the Gaussians are visualized by their contours of distribution density at 0.8 (the value of $e^{-\frac{1}{2} (\mathbf{x})^\mathrm{T} \boldsymbol{\Sigma}^{-1} (\mathbf{x})}$). 

\subsection{Radio Radiance Representation with CSI-Encoded Spherical Harmonics Function}

Given the explicit object geometry represented by 3D Gaussians with known positions, the next task is to model the RRF $\mathbf{c}(\mathbf{x}_{\text{obj}}, \mathbf{d})$. To this end, we adopt SH functions. As shown in Fig.~\ref{fig:5-RF-3DGS pipeline}~2.1, an SH function approximates a spherical function through a weighted sum of predefined SH basis functions. While the radio radiance $\mathbf{c}_{\mathbf{x}_{\text{obj}}}(\mathbf{d})$ at a given object point $\mathbf{x}_{\text{obj}}$ is also a spherical function, in RF-3DGS SH functions are not assigned to such explicit object points $\mathbf{x}_{\text{obj}}$, but rather to Gaussians. In particular, each SH function assigned to a Gaussian does not describe the radiance emitted from the center of that Gaussian. Instead, it represents the uniform yet view-dependent ``color'' (i.e., Spatial-CSI) of the entire Gaussian ellipsoid across different view directions, as illustrated in Fig.~\ref{fig:5-RF-3DGS pipeline}~2.3.
Therefore, for an object point $\mathbf{x}_{\text{obj}}$ in space, the corresponding radiance $\mathbf{c}_{\mathbf{x}_{\text{obj}}}(\mathbf{d})$ is represented as a density-weighted combination of the SH functions of all Gaussians that spatially overlap with $\mathbf{x}_{\text{obj}}$ (i.e., those with non-negligible Gaussian density at this point). Although this representation may appear complex, it enables highly efficient querying and reconstruction.
In implementation, the number of SH basis functions is determined by the SH degree. Higher degrees allow the representation of more complex and directional patterns. In our experiments at 60~GHz, where diffuse reflections dominate, an SH degree of 3 is sufficient. At lower frequencies, however, specular reflections become more significant, requiring higher SH degrees to capture sharper directional radiance variations.

Furthermore, since radio radiance involves multi-modal Spatial-CSI, we use parallel SH functions to separately represent each channel characteristic. Among them, path loss is the most critical one, as it indicates the importance of each MPC. In RF-3DGS, we use path loss as a basis for the encoding of other Spatial-CSI components, ensuring that important MPCs are more accurately modeled. For path loss, directly using its absolute value introduces a scale inconsistency. According to the Friis transmission equation:
\begin{equation}
    P_r = P_t \cdot \frac{G_t G_r \lambda^2}{(4\pi d)^2},
\end{equation}
where $P_r$ and $P_t$ are the received and transmitted power, $G_t$ and $G_r$ are the transmitter and receiver antenna gains, $\lambda$ is the wavelength, and $d$ is the full propagation distance. The received power decays proportionally to $1/d^2$.
However, referring back to Fig.~\ref{fig::3_RRF rendering} and Eq.~\eqref{eq::volume_rendering_eq}, it becomes clear that radiance field methods are inherently distance-invariant along a ray. This discrepancy arises because the radiance field method was originally designed to mimic visible light behavior. For visible light, when viewing an LED bulb from 1 meter and then 2 meters away, the perceived brightness does not drop by four times. This is explained by the human eye's response to light intensity, which follows a logarithmic-like curve and spans a dynamic range of approximately 100 dB~\cite{rose1948sensitivity}.
Motivated by this, we convert the pathloss spectrum to the dB scale to mimic such distance-independent decay. Notably, this lossy transformation is not a compromise of the radiance field method's limitations, but rather a way to take advantage of the visual intensity range. This transformation allows the spectrum to capture more details of the RRF,  as shown in Fig.~\ref{fig:8_bf_pattern}, where the dB scale spectrum more clearly describes the radio radiance field, at the cost of a few dB deviation in reconstruction. 

Then, the converted dB-scale path loss values are truncated and then normalized to pixel values in the range [0, 255]. The upper limit is set to the free space propagation loss at 0.5 meters to avoid unexpected high values in testing and online operation. The lower limit is the lowest sampled value in the training samples, usually the noise level. For improved visualization, most demonstrations in this paper apply a ``jet'' colormap to the path loss channel of the radio spatial spectrum (a mapping from gray image to the RGB image). Along with the encoding of the path loss value, the AoD and delay can be encoded using parallel channels. Specifically, AoD can be perfectly represented by our current RRF representation, as it only depends on the geometry of each ray. Therefore, we decompose it into two channels: one encoding the azimuth angle ($0^\circ$ to $360^\circ$) and the other encoding the zenith angle ($0^\circ$ to $180^\circ$). These two AoD angles of an MPC (pixel) are scaled to [0, 1] and then multiplied by the corresponding path loss pixel value. This concept is illustrated in Fig.~\ref{fig:5-RF-3DGS pipeline}~2.2, where the encoded spectrum is visualized as an RGB image, while two single-channel AoD spectra are shown as two grayscale images.

For the delay channel, we define ``delay" as the normalized delay of each MPC relative to the first-arrival MPC, ranging from 0 to 200 ns. MPCs exceeding 200 ns are discarded considering the indoor environment in this study. Although time of flight would be preferable, it is difficult to acquire in field measurements, and cannot be represented in the current RRF representation, as it is linearly related to the radio travel distance. The normalized delay alleviates this issue. For example, considering different Rx positions along a ray corresponding to the first-arrival MPC, the delay of later MPCs from other directions will slowly vary, which can be perfectly captured by the current RRF representation. However, once the different Rx positions are along a ray that does not correspond to the first-arrival MPC, the delays of other MPCs may deviate from the true delay. The larger the angle between the first-arrival MPC ray and the rays corresponding to later MPCs, the larger the deviations will be. However, such delay deviations are not significant, as shown in our later test of reconstructing delay spectra at unvisited locations. 

\subsection{Rendering RRF with the 3DGS Rasterization Pipeline}
Since the millions of 3D Gaussians collectively define a continuous density field, and each Gaussian is associated with an SH function representing its directional radio radiance, querying each discrete direction individually becomes computationally inefficient. Instead, we adopt a parallelized splatting approach, where each 3D Gaussian ellipsoid is projected onto the Rx's angular domain from the closer ones to the farther ones to form a 2D ellipse, as shown in Fig.~\ref{fig:5-RF-3DGS pipeline}~2.3. In this formulation, all query directions affected by this Gaussian can simultaneously accumulate its contributions to their final radio radiance.
Specifically, for each individual Gaussian, the splatting process is defined by the view transformation matrix $\mathbf{W}$ and the perspective projection matrix $\mathbf{P}$. To improve computational efficiency, we apply an affine Jacobian approximation $\mathbf{J}$ of the full perspective projection. Under this approximation, the projected 2D covariance matrix $\mathbf{\Sigma'}$ of the Gaussian is computed as:
\begin{equation}
    \mathbf{\Sigma'} = \mathbf{J} \mathbf{W} \mathbf{\Sigma} \mathbf{W}^\top \mathbf{J}^\top,
\end{equation}
where $\mathbf{\Sigma}$ is the 3D covariance matrix of the original Gaussian, and $\mathbf{\Sigma'}$ represents the covariance of the resulting 2D Gaussian after projection.
This affine approximation can be interpreted as each 3D Gaussian being first flattened to a 2D Gaussian by orthogonal projection along the camera depth axis, meaning that parts of the Gaussian do not appear larger as they get closer nor smaller as they move further away, which is the cost of the affine approximation. Then, the 2D Gaussians are normally projected onto the 2D plane, following the perspective projection principle. 

Beyond this, an efficient rasterization pipeline, well-suited for GPU computational structures, is employed. First, given the current rendering image plane and view frustum, the image plane is divided into $16~\times~16$ tiles, and the view frustum is divided accordingly. For each tile, only the Gaussians within the corresponding divided view frustum are selected (following a specific criterion) and sorted based on their depth to the plane. Finally, the computation within each tile consists of parallelized threads rendering the sorted Gaussians onto each pixel. Once a pixel is saturated with little transmittance, meaning that further Gaussians cannot affect this pixel, the thread rendering this pixel is terminated. Further details about the sorting process and GPU memory management for the millions of Gaussians can be found in~\cite{kerbl3Dgaussians}.

\begin{figure}[h] 
\vspace{-2mm}
\centering 
\includegraphics[width=0.4\textwidth]{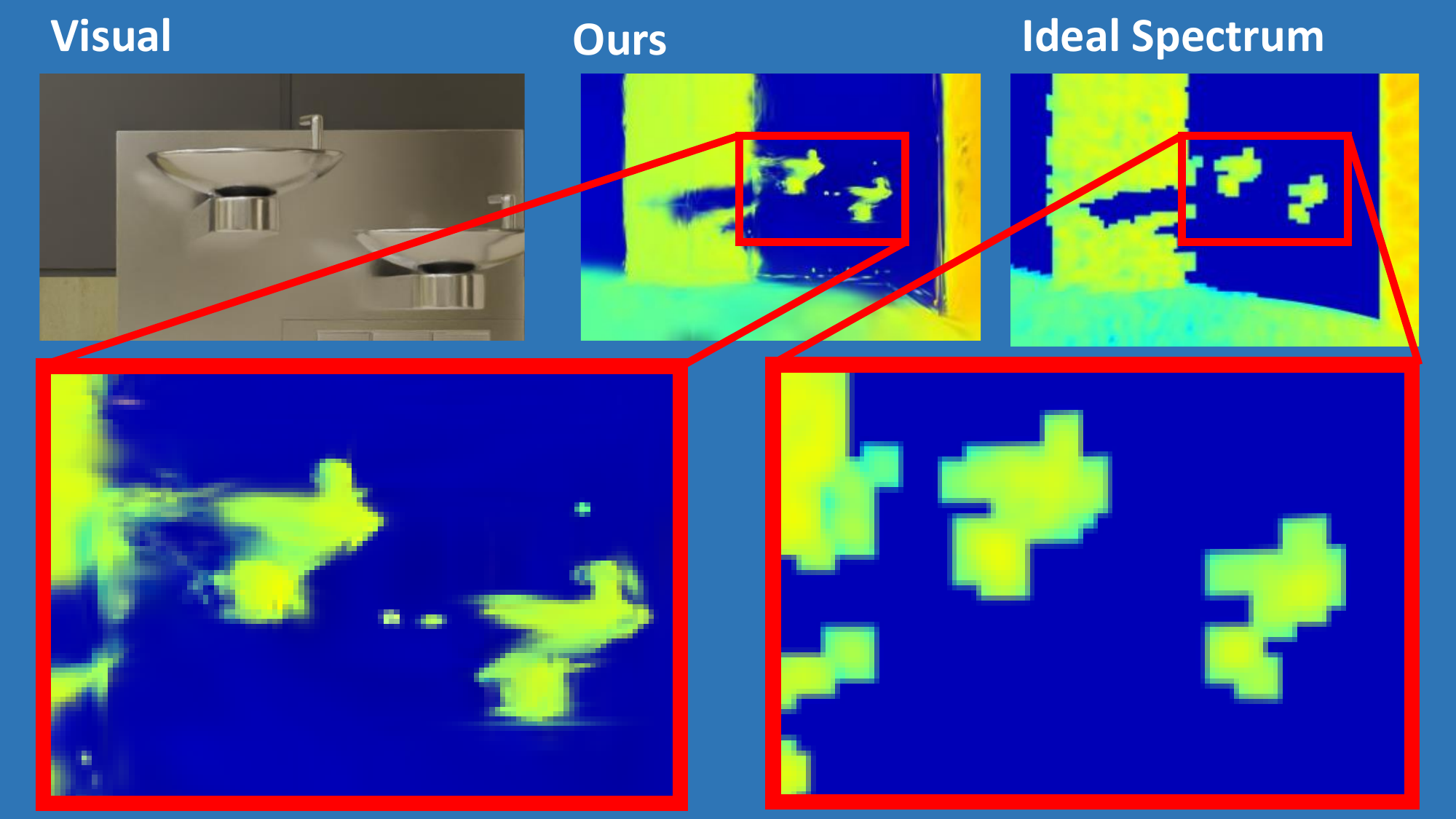} 
\vspace{-2mm}
\caption{Extrapolation ability of the proposed two-stage fusion training.
\textnormal{Both spectra are rendered using the same pinhole camera model. The ideal spectrum, serving as the training target, is limited by the maximum number of MPCs our RTX A6000 48~GB GPU can handle and the minimum array pattern lobe width needed for spectrum continuity.}} 
\vspace{-7mm}
\label{fig:extrapolation} 
\end{figure}

\subsection{Two-Stage Fusion Training}
Given the rendered spectra from the learnable RRF representation and the training spectra, the loss function used in this work is a combined loss function. This function is defined as the $L1$ loss plus a differential Structural Similarity Index Measure (D-SSIM) term:
\begin{equation}
    \mathcal{L} = (1 - \lambda) \mathcal{L}_1 + \lambda \mathcal{L}_{D-SSIM},
\end{equation}
where $\lambda$ is a predefined parameter, set to 0.2 in this work. The gradient descent then backpropagates this loss through the differentiable rasterization pipeline to the learnable parameters of each Gaussian.

A major challenge is that the low-quality radio spatial spectrum naturally lacks geometric information. From Fig.~\ref{fig:8_bf_pattern}, even the ideal spectrum can only be considered a low-quality ``photo", with other spectra offering even less geometric detail. In RF-3DGS, we assume access to the radio spatial spectra and the visual data, which serve as a low-cost supplement to geometric information. The proposed two-stage fusion training process is designed to extract geometric information from the visual data to enhance the RRF representation.

Before the training begins, if the visual data consists of photos or video without camera intrinsic and extrinsic parameters, an initialization process is required to estimate these parameters and generate initial point clouds. Otherwise, this process can be skipped, allowing training to start from random point clouds. 
Once initialization is complete, the first stage of training begins with a warm-up phase. During this phase, the Gaussians are trained using resolution-reduced images, enabling them to quickly capture global geometry rather than getting stuck on optimizing fine details. The image resolution is doubled twice, after 250 iterations and 500 iterations, until the original resolution is reached.

Following the warm-up, a periodic densification process refines the Gaussians every 100 iterations. In regions where the number of Gaussians is insufficient to represent intricate 3D geometry, the densification process splits large Gaussians and clones small Gaussians to create more learnable ``ellipsoids" in those regions. Gaussians with very small density values or those with excessively large footprints in the views are removed to prevent the ``floater" artifact problem, a common issue in radiance field reconstruction methods.
To manage the quantity of Gaussians, their basic density values $\alpha_g$ are reset towards zero every 3000 iterations. This reset allows key Gaussians to quickly regain their original density values during optimization, while abnormal or unnecessary Gaussians, whose density values increase slowly, are removed in subsequent densification steps.

After establishing a well-trained geometry representation, the second stage begins. In this stage, RF-3DGS uses the collected radio spatial spectra to train the CSI-encoded SH functions and the basic density of each Gaussian, while freezing the positions $\mathbf{x}$, rotation $\mathbf{q}$, and scaling $\mathbf{s_{scal}}$. This setup ensures that the geometry representation has the adaptive capability to handle inconsistencies between the visual and radio geometries (since certain materials may appear opaque in visual data but translucent under radio waves). After optimization, the new CSI-encoded SH functions can be interpreted as the objects being re-illuminated by the given Tx radio, with multiple channels corresponding to different Spatial-CSI properties.

The advantage of the proposed two-stage fusion training is not only that it requires low-cost visual data as a supplement, but also that it exhibits extrapolation capabilities, as shown in Fig.~\ref{fig:extrapolation}. This capability arises from the accurate visual geometric information, which provides the actual radiance source location, allowing vague or interfered radio spectra to be mapped to the correct locations rather than being assigned to vague floaters, as seen in \textit{NeRF$^2$}.

\section{Digital Twin Framework}

While RF-3DGS largely addresses the core challenges in RRF reconstruction, the issue of low-quality radio spatial spectra remains only partially resolved. In realistic environments, generating radio spectra close to the simulated ideal or even MVDR cases is still prohibitively expensive. Therefore, it is crucial to evaluate how the degradation of training data affects RF-3DGS performance in practical scenarios.
To this end, we propose a digital twin framework that serves three purposes: (1) to generate ideal radio spectra for developing and validating RF-3DGS in a mostly controlled setting, (2) to evaluate the effects of spectra generated by various practical ASP algorithms and systems, and (3) to investigate how RRFs initialized from simulations can be calibrated using real-world field measurements.

\begin{figure}[h]
    \centering
 
    \includegraphics[width=0.4\textwidth]{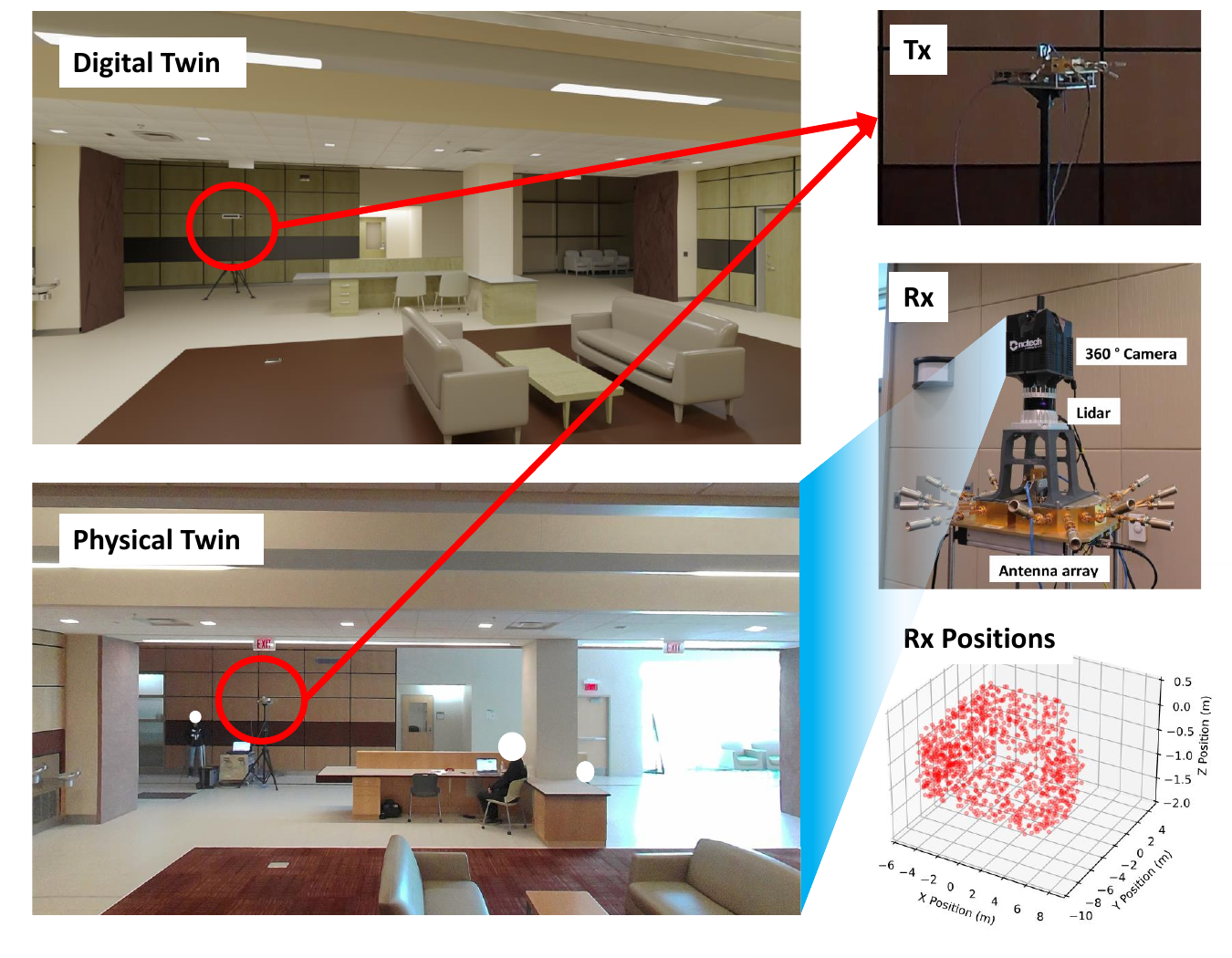}
    \vspace{-3mm}
    \caption{Dataset overview.}
    \vspace{-3mm}
    \label{fig::Dataset overview}
\end{figure}

\begin{figure*}[ht]
  
    \centering
    \includegraphics[width=0.9\textwidth]{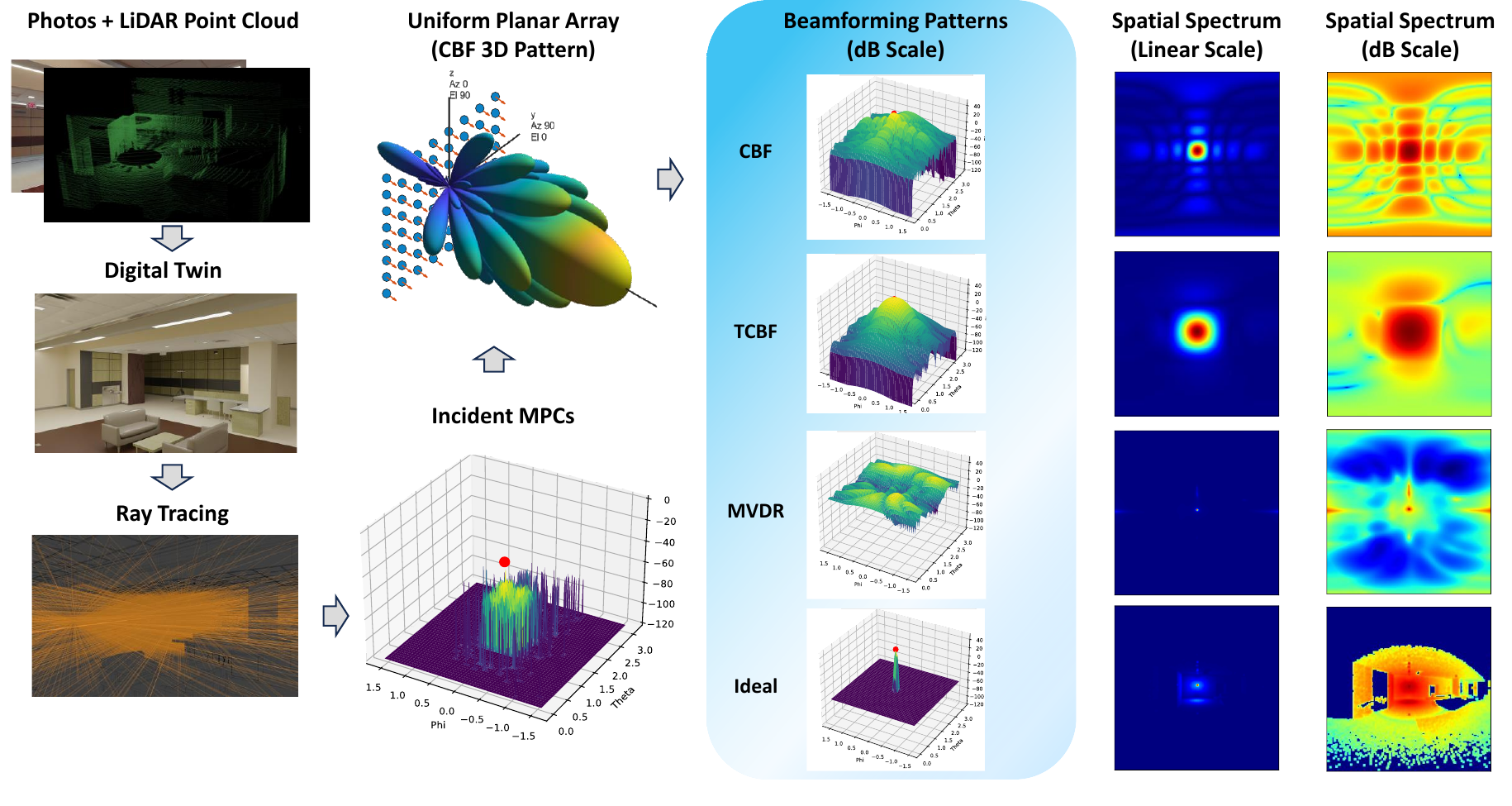}
    \vspace{-2mm}
    \caption{Full radio simulation pipeline with various beamforming patterns and their corresponding spectra.}
    \vspace{-3mm}
    \label{fig:8_bf_pattern}
\end{figure*}

\textbf{Physical twin measurement:}  
As illustrated in Fig.~\ref{fig::Dataset overview} and Fig.~\ref{fig:8_bf_pattern}, we construct a digital twin of a real-world lobby environment as the playground. The field measurement campaign, conducted by the National Institute of Standards and Technology (NIST), covers a 14~m $\times$ 15~m lobby space. Measurements were taken using horn antenna arrays at a fixed Tx and a mobile Rx along predefined routes around the lobby.
The raw 60~GHz measurements are processed using the Space-Alternating Generalized Expectation-maximization (SAGE) algorithm, which extracts MPCs with detailed Spatial-CSI between the Tx and each Rx location. Further details on this system setup and signal processing can be found in~\cite{NIST_data,NIST_SAGE}. However, a major limitation of SAGE is that the extracted MPCs tend to be sparse and angularly inconsistent among different views. For instance, in the scenario shown in Fig.~\ref{fig::4_RRF reconstruction}, such inconsistency can prevent two red rays from intersecting at the same surface point in 3D space. In this context, if these MPCs are directly mapped to individual pixels in the spatial spectrum, the resulting training spectrum may conflict during training, leading to reconstruction failure.

\begin{figure}[h]
    \centering
    \includegraphics[width=0.32\textwidth]{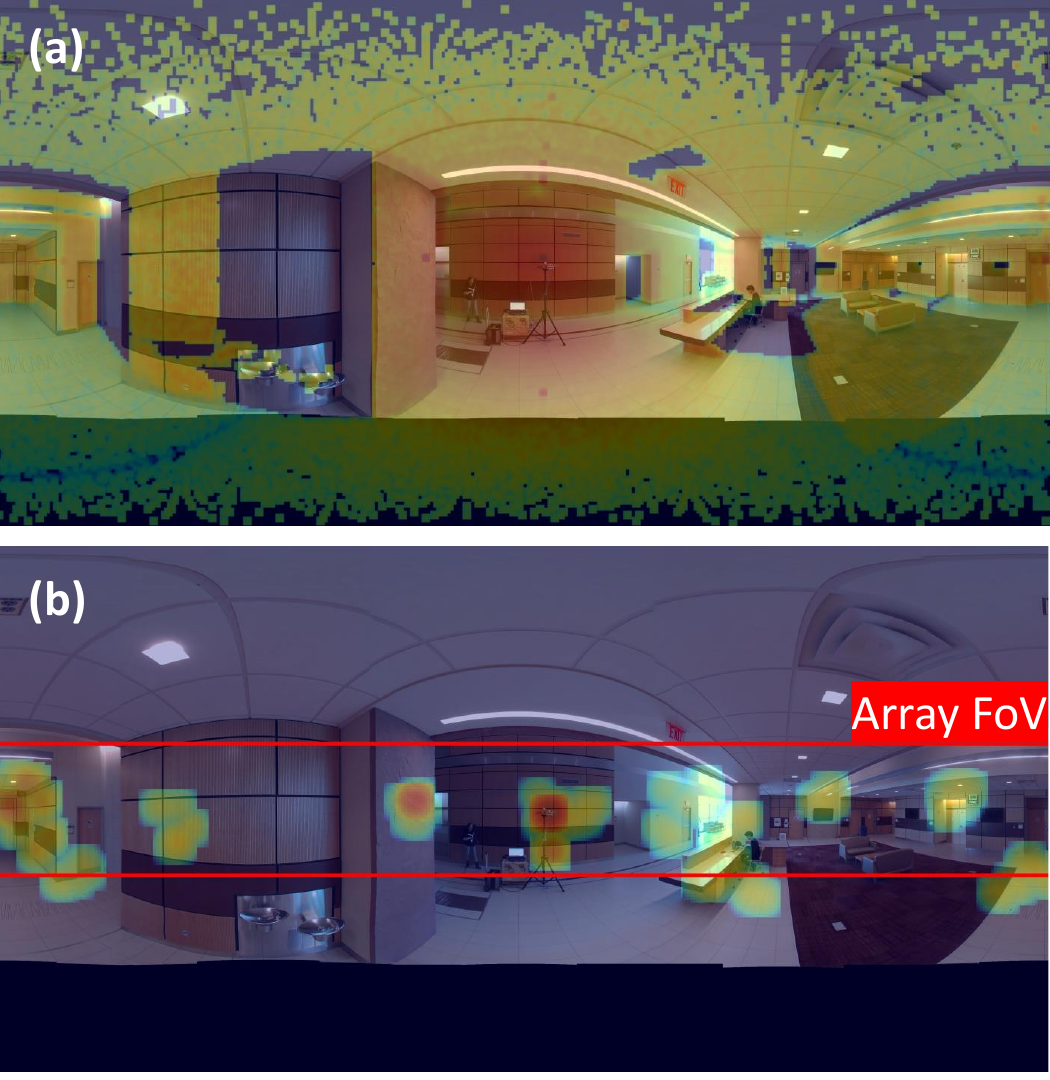}
    \vspace{-3mm}
    \caption{(a) Digital replica and (b) processed field measurements~\cite{NIST_data}.}
    \label{fig::physical twin measurement twin}
    \vspace{-5mm}
\end{figure}

To address this, we introduce angular domain ambiguity similar to that used in geometry modeling. Specifically, we spread the extracted MPCs to form more continuous signal patterns, leading to the SAGE spectra as shown in Fig.~\ref{fig::physical twin measurement twin}(b). The two red lines represent the FoV of the measurement arrays. This ambiguity preserves training consistency across views.
Furthermore, to support the digital twin environment, we collect rich multi-modal data around the lobby, including approximately 1000 co-located photographs and LiDAR scans. 

\label{generating spectrum}

\textbf{Digital Replica Establishment:}  
Based on the collected multi-modal data in the lobby, we first fuse LiDAR scans and image features into a global point cloud, from which a coarse single environment mesh is extracted. This mesh is then manually refined into a high-resolution model with accurate geometry. We assign visual textures to each surface and semantic labels to facilitate later RF parameter assignment, resulting in the digital replica shown in Fig.~\ref{fig:8_bf_pattern}. This digital replica is imported into the Simulator Sionna, an open-source communication simulator developed by NVIDIA, which is equipped with the Sionna RT~\cite{hoydis2023sionna} ray tracer and supports GPU acceleration. Within the simulator, we further assign EM material properties and their corresponding reflection models to each semantic label, and consequently to each surface mesh in the digital replica. Then, we replicate the field measurement setup by fixing the Tx at the same position used in the physical measurement. The Rx positions are randomly distributed throughout the environment to capture a wide range of propagation conditions.
At each Rx, the ray tracing simulation generates MPCs distributed in its angular domain, received by a simulated UPA. These simulated responses at the UPA elements are subsequently processed using different ASP algorithms to generate radio spatial spectra. %Further details regarding simulation parameters and processing configurations are available in our released codebase.

\textbf{Evaluation of ASP Algorithms and Systems:}  
We evaluate four types of radio spectra, each generated using a distinct ASP setup: (1) Conventional Beamforming, (2) Tapered Conventional Beamforming (TCBF), (3) MVDR, and (4) the ideal spectrum. The simulated environment assumes a highly diffuse propagation condition at 60~GHz, resulting in more than 300,000 MPCs per Tx-Rx pair. Due to computational constraints, we limit the simulated array to an $8 \times 8$ uniform planar array composed of patch antenna elements with half-wavelength inter-element spacing, and use this as the Rx device for the first three setups. For the ideal spectrum, we use a synthesized array gain pattern designed to minimize the influence of data imperfections.

For ASP algorithms, the CBF, also known as the Bartlett beamformer~\cite{bartlett1950periodogram}, is a delay-and-sum beamforming technique. It adjusts the phase shifts of each array element to align their wavefronts in the desired scanning direction, only compensates for phase differences, and thus leads to strong side lobe levels and interference in the final generated spectrum. 
To suppress side lobes, we employ TCBF, which applies a Hann window~\cite{van2002optimumarray} to the beamforming vectors. While this reduces side-lobe levels, it leads to a wider main lobe and still fails to provide sufficient interference suppression when the spectrum is visualized on the dB scale.
To address this, we apply MVDR beamforming~\cite{capon1969high}, also known as the Capon beamformer. MVDR optimizes the beam pattern to maximize the signal-to-interference-plus-noise ratio (SINR), adaptively nulling interference while preserving acceptable gain in the scanning direction. Unlike CBF, which only maximizes the gain on the scanning direction, MVDR solves an optimization problem using the received responses on each array element, allowing it to steer nulls toward actual interference directions and largely suppress them.

This trade-off between main-lobe performance (sharpness and gain) and side-lobe suppression is illustrated across the methods: CBF maximizes main-lobe gain but suffers from high interference; TCBF suppresses all side lobes but broadens the main lobe; MVDR finds a balance by adaptively only suppressing the actual interference while maintaining good main lobe performance. As we will demonstrate in later sections, CBF and TCBF fail to support accurate RRF reconstruction, while MVDR delivers significantly improved reconstruction quality. However, this comes at the cost of increased hardware complexity—MVDR requires a fully digital beamforming system or a sophisticated setup capable of acquiring synchronized responses at all array elements, in contrast to CBF and TCBF, which are compatible with fully analog beamforming architectures.
Lastly, we simulate the ideal spectra with a synthetic beam pattern that emulates the pattern from a $64 \times 64$ UPA with side-lobe suppression. This spectrum serves as the upper bound for achievable quality in our current digital twin framework.

\section{Experimental Result and Evaluation}

In this section, to evaluate the performance of RF-3DGS, we designed several experiments. First, we directly compared the performance of RRF reconstruction with other methods. Next, to assess its generalizability, we tested how different types of spectra affect the performance of RF-3DGS. To evaluate RF-3DGS's effectiveness in supporting communication systems, we conducted experiments to test its ability to predict the CBF steering vector. Finally, as RF-3DGS reconstructs multi-dimensional CSI, we tested the accuracy of the reconstructed AoD and delay channels.

\textbf{Performance of Radio Radiance Field Reconstruction:}  
This experiment focuses on evaluating the RRF reconstruction performance of RF-3DGS compared to other methods. The dataset consists of 800 samples at randomly selected Rx positions, split into 640 for training and 160 for testing. To analyze performance under different sample densities, we randomly sample subsets of the training data with sizes $\{10, 25, 50, 100, 128, 256, 384, 512, 640\}$. To avoid the influence of ASP algorithms, all spectra used in this experiment are generated using the ideal spectra setup.
For each method under evaluation, we train them using spectra from the selected training subset. The trained models are then used to synthesize spectra at the test poses, which are compared to ground-truth ideal spectra generated at the same poses with the test samples. 
To evaluate the similarity between the synthesized spectra and the ground truth, we adopt three metrics: Peak Signal-to-Noise Ratio (PSNR), Structural Similarity Index Measure (SSIM), and Learned Perceptual Image Patch Similarity (LPIPS). (Despite its name, PSNR is computed from the mean squared error (MSE) between all pixel values, averaged over the image size, and then transformed into a logarithmic scale for more intuitive interpretation.)

\begin{figure}[h]
    \vspace{-2mm}
    \centering
    \includegraphics[width=0.45\textwidth]{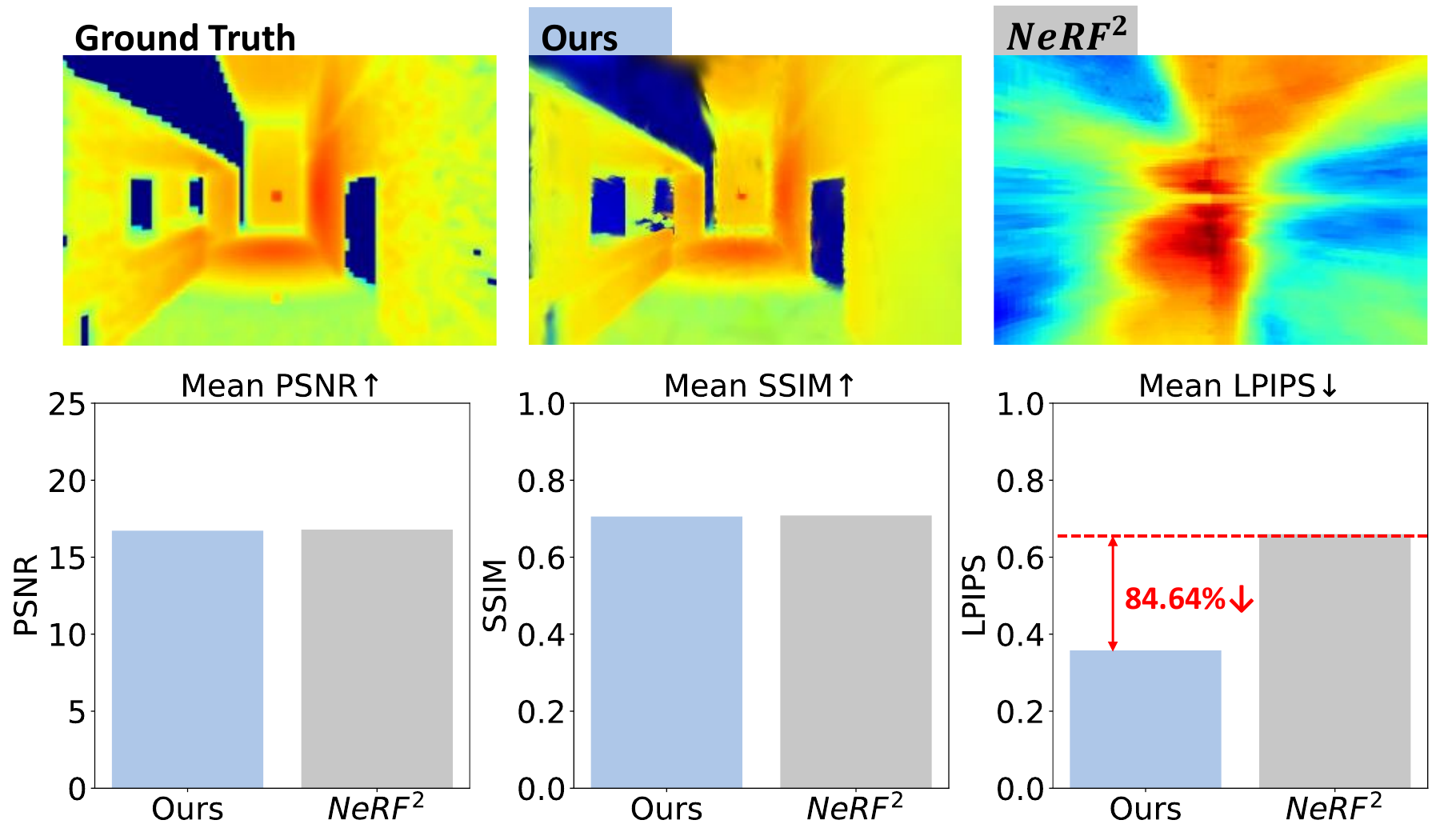}
    \vspace{-2mm}
    \caption{Limitation of statistical metrics.}
    \label{fig::metric_degradation}
    \vspace{-2mm}
\end{figure}

It is important to note that only high-level perceptual metrics, such as LPIPS, effectively capture visual and structural differences in reconstructed spectra. Traditional statistical metrics, such as the PSNR and SSIM, fall short. This issue has been widely discussed in the CV  domain~\cite{zhang2018unreasonable,lin2011perceptualmetric}. As illustrated in Fig.~\ref{fig::metric_degradation}, even though the spectrum predicted by our method is visibly of higher quality, the PSNR and SSIM yield similar values for both spectra, while only LPIPS reflects the difference accurately.
The main reason for this metric failure is that our scenario differs significantly from visual tasks. In typical visual tasks, like evaluating compression or super-resolution, outputs are usually close to the perfect ground truth images, resulting in high metric scores (PSNR~$>$~35 dB, SSIM~$>$~0.9, LPIPS~$<$~0.05), where errors are mostly pixel-wise.

However, RRF reconstruction is considerably more challenging. Here, PSNR is often around 15~dB, SSIM near 0.5, and LPIPS around 0.4, even for ideal conditions.
Additionally, the ``ideal" spectra ground truth for training and testing also contains distortions and noise. These factors reduce the metric discriminative power, as critical errors, such as the loss of geometric structure, are obscured by less significant noisy errors. This issue is particularly severe for PSNR and SSIM, which heavily rely on statistical features, unlike LPIPS, which utilizes convolutional neural networks to extract deep features for similarity evaluation.

\begin{figure}[h]
    \centering
    \includegraphics[width=0.48\textwidth]{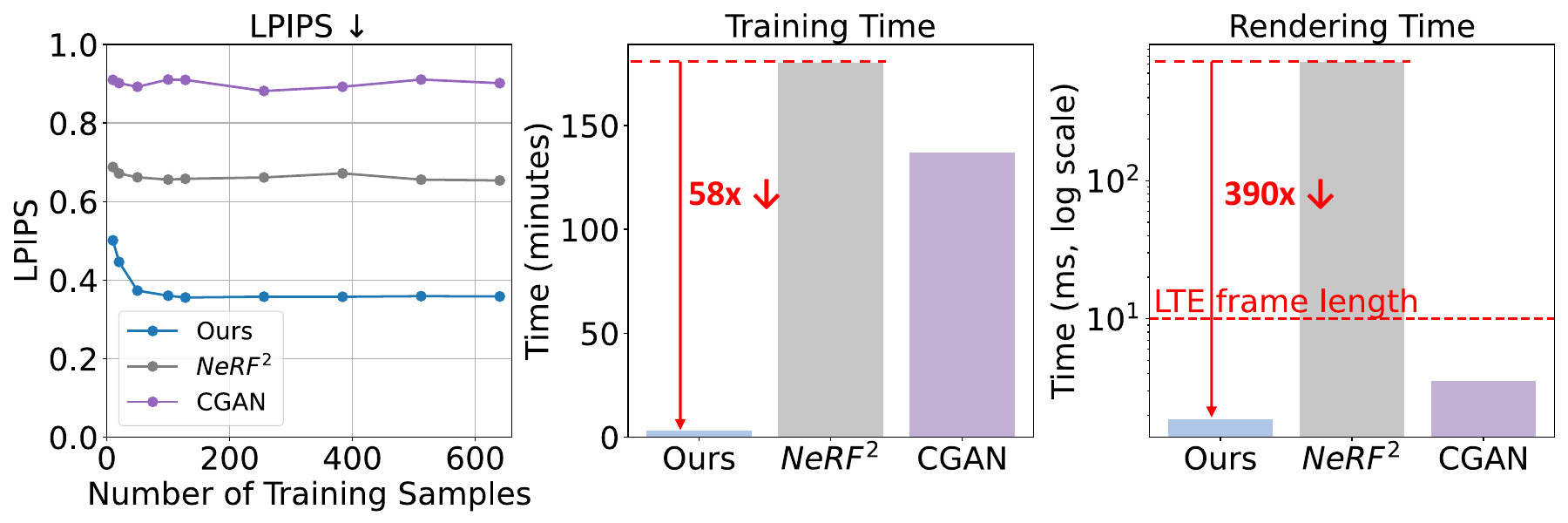}
    \vspace{-2mm}
    \caption{Comparison of RRF reconstruction performance.}
    \label{fig::RF-3DGS&Nerf2 on ideal spec}
    \vspace{-2mm}
\end{figure}

With LPIPS as the primary metric, we compare our method, RF-3DGS, with the Conditional GAN (CGAN)~\cite{mirza2014conditionalgan}, representing black-box neural network approaches, and the SOTA \textit{NeRF$^2$}. The results are presented in Fig.~\ref{fig::RF-3DGS&Nerf2 on ideal spec}.

RF-3DGS demonstrates significantly better LPIPS scores, requires orders of magnitude shorter training times, and offers rendering speeds faster than Long Term Evolution (LTE) frame rates. Additionally, RF-3DGS achieves high reconstruction quality with very sparse training samples, with rendering quality substantially degrading only when the sample count drops below approximately 20, depending on the uniformity of the sample distribution. For evaluating other parameters, we utilized the full training set to analyze their impact.
In contrast, \textit{NeRF$^2$} struggles to capture detailed scene information, resulting in degraded LPIPS performance. As illustrated in Fig.~\ref{fig::metric_degradation}, \textit{NeRF$^2$} can only reconstruct large, illuminated floaters with limited details. Moreover, its training process takes approximately 3 hours, with each rendering requiring around one second.
For CGAN, the generator synthesizes images while the discriminator evaluates their fidelity, guiding the generator's optimization. The challenge lies in our sparse training input, where test inputs differ significantly from the training data. Consequently, the generator struggles to learn the mapping between Rx poses and target images, and the discriminator fails to provide useful gradients, resulting in poor overall performance.

\begin{figure}[h]
    \centering
    \includegraphics[width=0.45\textwidth]{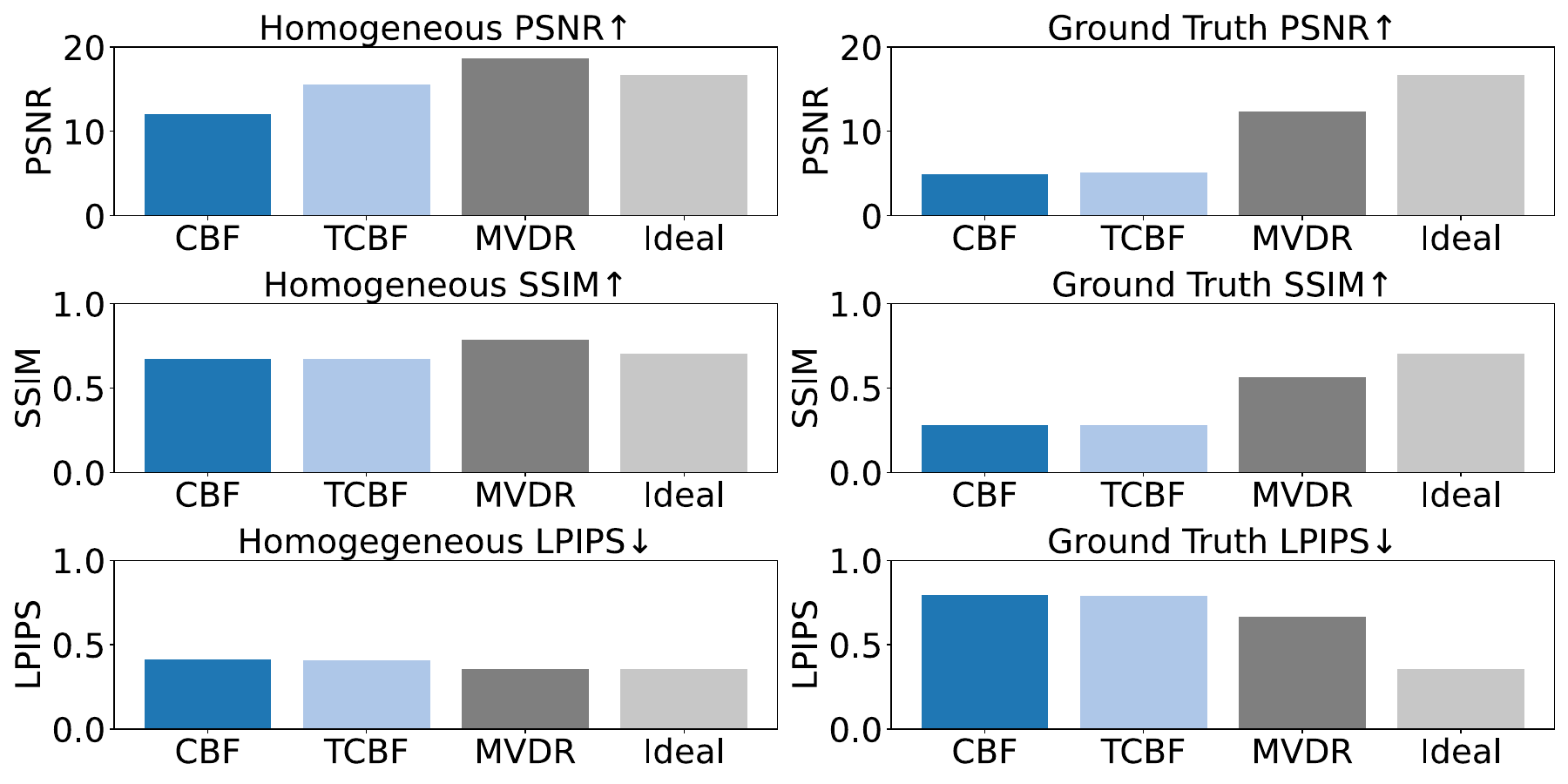}
    \vspace{-2mm}
    \caption{Impacts of different spectra.}
    \vspace{-2mm}
    \label{fig::spectrum_degradation}
\end{figure}

\textbf{Impacts of Different Spectra:} To further investigate the effect of different types of input spectra on reconstruction performance, we additionally trained RF-3DGS using spectra generated by CBF, TCBF, and MVDR setups. Each trained model was similarly evaluated at test poses but using two types of ground truth: Homogeneous spectrum ground truth (the same type of spectrum used for training each model) and ideal spectrum ground truth (ideal spectra for all models). These two tests serve distinct purposes: first, homogeneous spectrum testing evaluates how spectra interference and inconsistency impact RF-3DGS's reconstruction ability; second, ground truth testing assesses how these spectra degrade overall reconstruction quality compared to the most ideal case. Fig.~\ref{fig::spectrum_degradation} provides a detailed comparison.

From the comparison, it is clear that the CBF and TCBF spectra impair both reconstruction ability and quality, evident from their poor performance across all metrics. The degradation in reconstruction quality is even more pronounced, with PSNR dropping to 5 dB. This suggests that for CBF and TCBF spectra, strong interference and lack of geometric information severely limit the effectiveness of radiance field reconstruction in large environments.
In contrast, the MVDR spectrum group exhibits much higher performance than CBF and TCBF in both testing scenarios. More interestingly, in homogeneous spectrum testing, the MVDR spectrum group even results in higher PSNR and SSIM compared to the ideal spectrum group, which may appear counter-intuitive given the ideal spectrum’s higher quality and lower interference. However, as shown in Fig.~\ref{fig:extrapolation}, our method has an extrapolation ability that provides more details than the ideal spectrum. This extrapolation results in lower performance in statistical metrics like PSNR and SSIM but yields similar evaluations from LPIPS, further validating the effectiveness of the LPIPS metric in our tasks.

In summary, these comparisons demonstrate that RF-3DGS achieves significantly higher radio radiance field reconstruction quality, faster training speed, and faster rendering speed compared to other methods. Additionally, RF-3DGS requires only tens of samples to reconstruct the radiance field across a large lobby while providing an explicit geometric representation. These features and high performance underscore the potential of RF-3DGS in future 6G network applications.

\begin{figure}[h]
    \centering
    \includegraphics[width=0.35\textwidth]{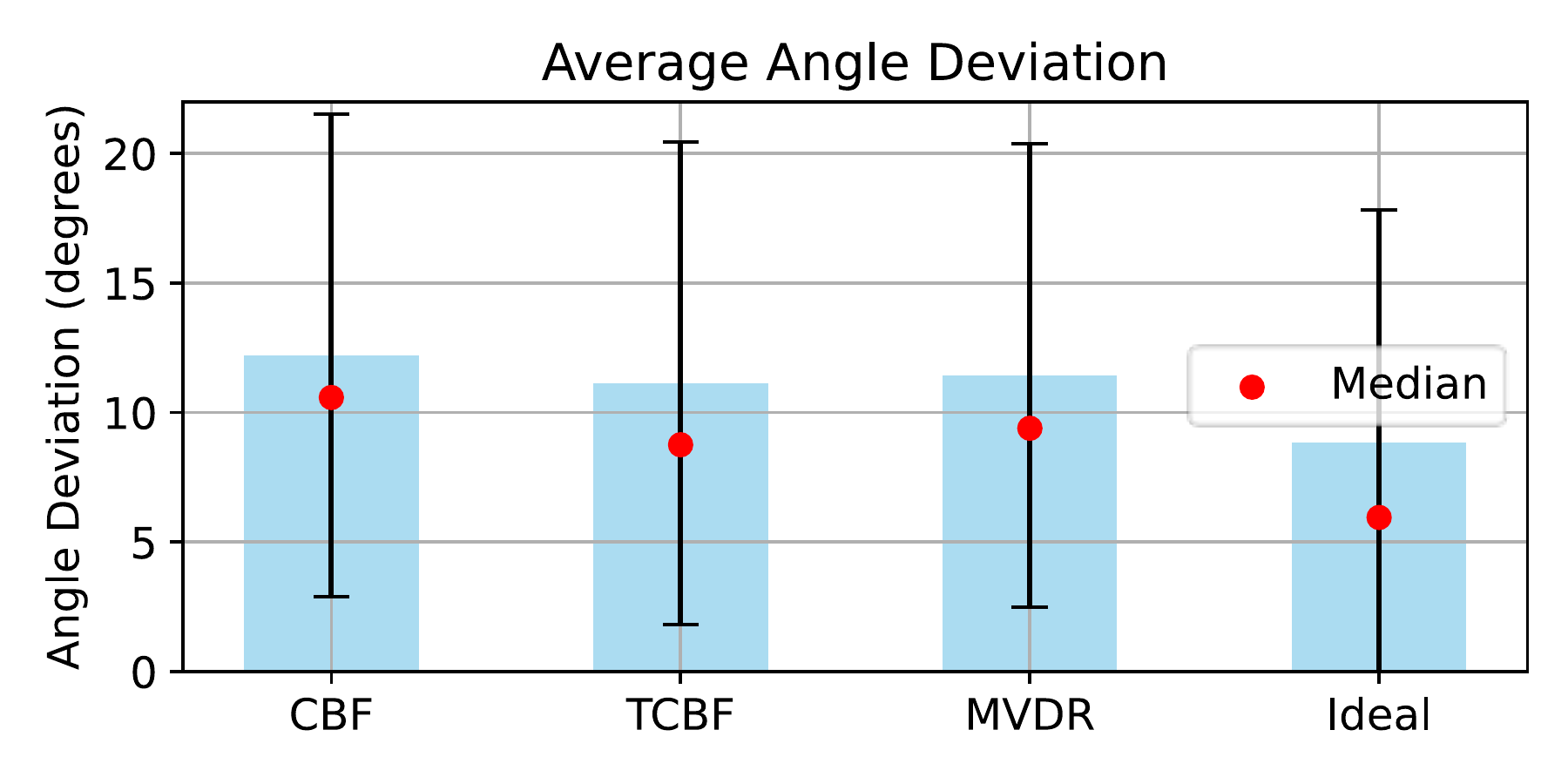}
    \vspace{-3mm}
    \caption{Angle deviation of CBF guided by spectra.}
   
    \label{fig::CBF receiving power}
\end{figure}

\textbf{Supporting Wireless Communication:} To evaluate the effectiveness of RF-3DGS in wireless communication applications, we consider a practical scenario where the Rx-side path loss spectrum is used to guide the Rx array in performing angular domain CBF. In this setting, the AoA corresponding to the strongest path is of primary importance, as it determines the beam steering direction.
In particular, we train RF-3DGS models using the four types of spectra (Ideal, MVDR, TCBF, and CBF), and then predict the CBF main lobe direction at unseen test positions. The ground truth is defined as the AoA of the strongest MPC in the ideal spectrum generated from the sample at each test location.
We report the average angular deviation between the predicted main lobe directions and the ground truth across the test set. This deviation reflects the degree of beam misalignment, and in practical systems, corresponds to additional overhead required for beam calibration.
As shown in Fig.~\ref{fig::CBF receiving power}, although the CBF and TCBF spectra fail to accurately capture the underlying scene geometry, they are still able to support basic beamforming operations with moderate performance loss. 
\vspace{-2mm}

\begin{figure}[h]
    \centering
    \includegraphics[width=0.45\textwidth]{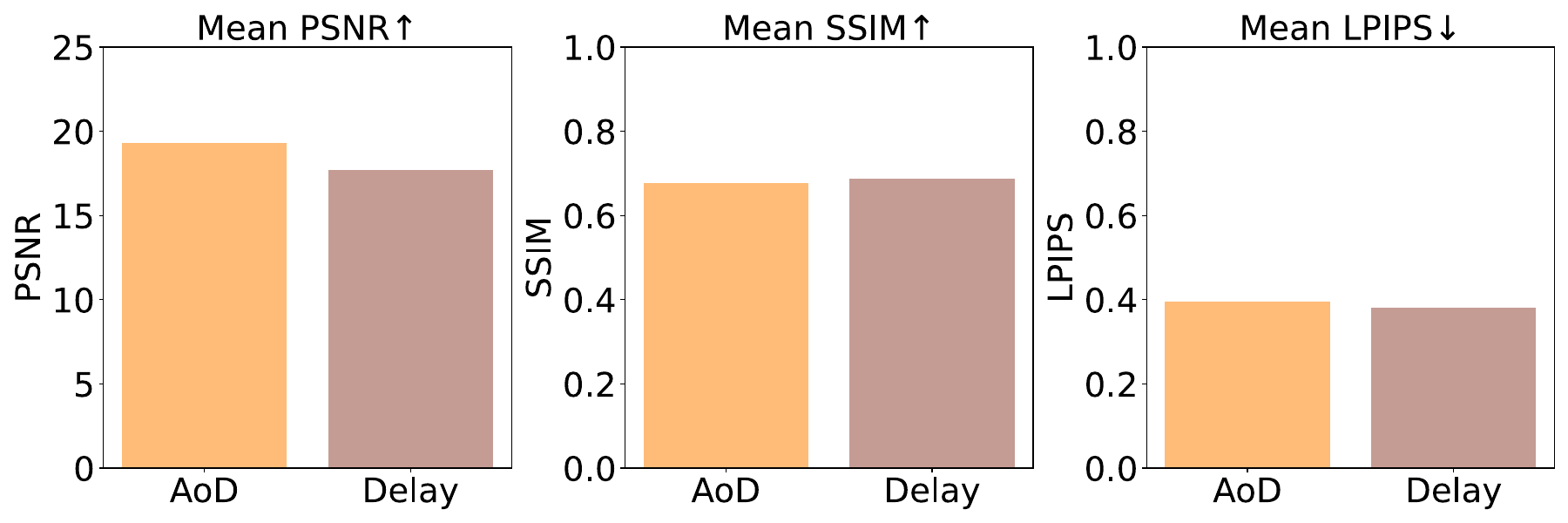}
    \vspace{-0.2cm}
    \caption{Accuracy of encoded Spatial-CSI spectra.}
    \vspace{-0.2cm}
    \label{fig::CSI spectrum accuracy}
\end{figure}

\textbf{Spatial-CSI Prediction:}  
Another key feature of RF-3DGS is its capability to encode Spatial-CSI. In this paper, we test its capability to represent the AoD and delay, which are critical in advanced wireless communication and sensing applications.
As shown in Fig.~\ref{fig::CSI spectrum accuracy}, the results indicate the ability of RF-3DGS to reconstruct Rx-side AoD spectra and delay spectra. We also present examples of the decoded Spatial-CSI spectra in Fig.~\ref{fig:5-RF-3DGS pipeline}. For the delay channel, the example demonstrates that the delay increases from the near reflection point to the far end, similar to a depth image. In the AoD-azimuth channel, although the angles vary according to physical laws, a sharp edge corresponding to the transition between 0 and 360$^{\circ}$ is visible. A more appropriate approach would involve using a 3D unit vector to represent such AoD information; however, in this paper, we use these two angles for a more intuitive demonstration.

\section{Field Study: A Digital Radio Twin}
In previous experiments, our testing relied on simulation datasets, which raised concerns about potential discrepancies between our digital twin and its physical counterpart. In this section, we demonstrate the fidelity of our digital twin framework and highlight its support for advanced applications such as ISAC.

\begin{figure}[ht]
    \centering
    \includegraphics[width=0.45\textwidth]{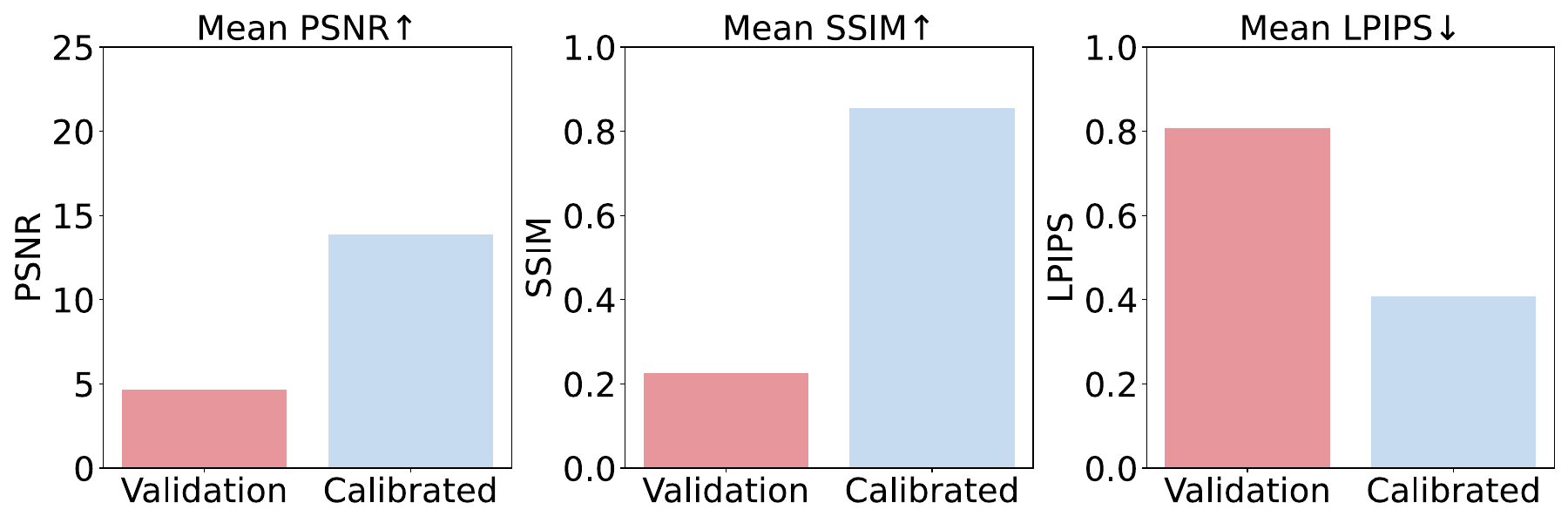}
    \vspace{-2mm}
    \caption{Calibration with field measurement.}
    \vspace{-2mm}
    \label{fig::field_mearsure_vali}
\end{figure}

\textbf{Calibrating the Radiance Field:}  
To validate the fidelity of our constructed digital replica, we compare the ideal spectra generated from simulation with the real-world SAGE spectra extracted from field measurements. For direct comparison, we overlay both spectra with equirectangular visual photographs captured from corresponding Rx poses. As illustrated in Fig.~\ref{fig::physical twin measurement twin}, the simulated ideal spectra show strong alignment with the dominant MPCs identified in the SAGE spectra, as well as with the scene geometry. We further use the SAGE spectra to numerically validate and calibrate the reconstructed RRF, mimicking real-world online system operations where the physical system exhibits imperfections. In such scenarios, the efficiency and extrapolation capability of RF-3DGS become particularly advantageous. The model can be fine-tuned using the real SAGE spectra with just 3 minutes of re-training, resulting in an updated radiance field that achieves high performance on test metrics, as shown in Fig.~\ref{fig::field_mearsure_vali}.

\begin{figure}[h]
    \centering
    \includegraphics[width=0.5\textwidth]{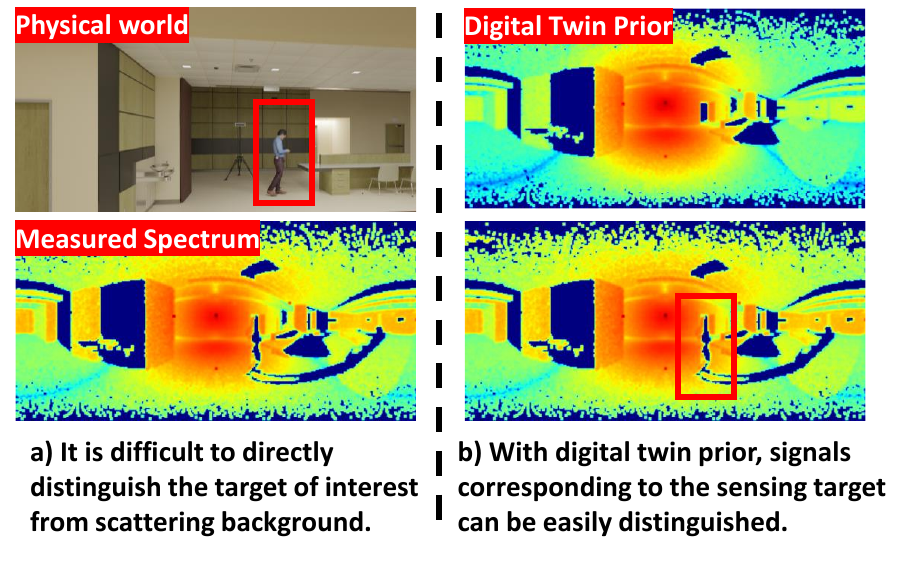}
     \vspace{-8mm}
    \caption{ISAC scenario demonstration. }
     \vspace{-2mm}
    \label{fig::ISAC scenario demonstration}
\end{figure}

\textbf{An ISAC Scenario Demonstration with Digital Twin:}  
Given the strong reconstruction capability of RF-3DGS and its ability to integrate real-time wireless communication feedback, the framework is naturally well-suited for ISAC applications~\cite{isac}. Fig.~\ref{fig::ISAC scenario demonstration} illustrates a representative ISAC scenario implemented within the digital twin.
In this example, a person enters the scene and becomes the target of interest for sensing. Meanwhile, a device equipped with an antenna array observes the environment and captures a radio spatial spectrum. However, the raw spectrum contains a large number of MPCs originating from the scattering background. Traditional sensing approaches often struggle to distinguish the signal component associated with the sensing target~\cite{wifisen} from such background.
In contrast, as shown in Fig.~\ref{fig::ISAC scenario demonstration}(b), the digital twin provides a learned prior of the scattering background, and makes the MPCs introduced by the sensing target easy to identify. These signal components can then be extracted for further sensing tasks such as localization, gesture recognition, or human activity analysis.
\vspace{-2mm}

\section{Conclusions}
In this paper, we proposed RF-3DGS, a fast and efficient method for radio radiance field reconstruction that achieves high performance in both training efficiency and reconstruction quality.  
Despite its effectiveness, several challenges remain.  
First, acquiring accurate radio spatial spectra remains costly and impractical in many real-world scenarios.  
Second, in more dynamic device-to-device communication where the Tx is also mobile, accommodating such variations poses a significant challenge.  
Moreover, addressing deviations in the RRF representation requires further research into novel representation designs and rendering pipelines.  
Finally, expanding the practical applications of RF-3DGS, particularly in emerging 6G scenarios such as cell-free massive MIMO and ISAC, represents a promising direction for future work.

\bibliographystyle{ieeetr}
\bibliography{sample-base_modified}

\end{document}